\documentclass[12pt]{iopart}
\usepackage[dvips]{graphicx}
\usepackage{epsfig}
\usepackage{subfigure}
\usepackage{epsfig}
\usepackage{color}
\newcommand{\half}{\mbox{$\frac{1}{2}$}}

\newcommand{\lb}{\label}

\newcommand{\bss}{\mathbf{s}}

\begin{document}
\title[The Effect of Nonstationarity on Models Inferred from Neural Data]{The Effect of Nonstationarity on Models Inferred from Neural Data}
\author{Joanna Tyrcha}
\address{Department of Mathematical Statistics, Stockholm University, 10691 Stockholm, Sweden}
\author{Yasser Roudi}
\address{Kavli Institute for Systems Neuroscience, NTNU, 7010 Trondheim, Norway and
Nordita, Stockholm University and the Royal Institute of Technology, 106 91 Stockholm, Sweden}
\author{Matteo Marsili}
\address{ICTP, Strada Costiera 11, 34014, Trieste, Italy}
\author{John Hertz}
\address{Nordita, Stockholm University and the Royal Institute of Technology, 106 91 Stockholm, Sweden and
the Niels Bohr Institute, 2100 Copenhagen, Denmark} 

\begin{abstract} 
Neurons subject to a common non-stationary input may exhibit a correlated firing behavior.
Correlations in the statistics of neural spike trains also arise as the effect of
interaction between neurons. Here we show that these two situations can be distinguished,
with machine learning techniques, provided the data are rich enough. In order to do this,
we study the problem of inferring a kinetic Ising model, stationary or nonstationary, from
the available data. We apply the inference procedure to two data sets: one from salamander
retinal ganglion cells and the other from a realistic computational cortical network
model. We show that many aspects of the concerted activity of the salamander retinal
neurons can be traced simply to the external input.  A model of non-interacting neurons
subject to a non-stationary external field outperforms a model with stationary input with
couplings between neurons, even accounting for the differences in the number of model
parameters. When couplings are added to the non-stationary model, for the retinal data,
little is gained: the inferred couplings are generally not significant. Likewise, the
distribution of the sizes of sets of neurons that spike simultaneously and the frequency
of spike patterns as function of their rank (Zipf plots) are well-explained by an
independent-neuron model with time-dependent external input, and adding connections to
such a model does not offer significant improvement.  For the cortical model data, robust
couplings, well correlated with the real connections, can be inferred using the
non-stationary model.  Adding connections to this model slightly improves the agreement
with the data for the probability of synchronous spikes but hardly affects the Zipf plot.
\end{abstract}
\maketitle
\section{Introduction} 
A significant amount of work in recent years has been devoted to finding simple
statistical models of data recorded from biological networks
\cite{Schneidman06,Shlens06,Lezonetal06,Yu08,Cocco09,Roudi09,Roudi09-2,WeigtPNAS2009}.
Using the output of recordings from many neurons or many genes, this body of work aims at
better understanding the collective behavior of elements of a biological network and
gaining insight into the relationship between network connectivity and correlations
between these elements.

The pattern of connectivity in a biological network, however, is not the only source of
correlations.  Another important factor in shaping large scale concerted activity is the
effect of time-varying external input.  This effect is often neglected in empirical
studies, because rich data are needed to resolve the time dependence.  Such input can
induce apparent correlations that, if not taken into account properly, can lead to
artifacts. For instance, from the correlated activity of two neurons, one might infer that
there is a (direct or indirect) connection between them, while in reality the observed
correlation could be due to correlated external input that they receive.  For sensory
system neurons, such correlations are frequently called ``stimulus-induced''.   Of course,
if the time dependence of their firing rates is known, these apparent correlations can be
removed, but commonly in experiments they are not known and therefore simply assumed
constant in time. When trying to learn something about a network by fitting a statistical
model to it, it is therefore important to separate the aspects of the data mediated by
internal network circuitry from those which are simply reflections of time-dependent
external input.

To better appreciate the importance of this point, one can draw an analogy to a spin
system with the spins mostly ordered in one direction. For this system, the order can be
due to the interactions between spins being strong, leading them to align in a particular
direction (as in ferromagnets). Alternatively, this order can just be due to the presence
of an external field aligning the spins. Needless to say, the two scenarios lead to
substantially different pictures of the system. For biological systems, which in most
cases are subject to time varying and correlated input from the external world, it would
thus make sense to focus on statistical models that allow such nonstationarity and
correlations and see what connections and external fields are inferred when no assumptions
about the temporal dynamics of the input are forced upon the model. Using equilibrium
models, a number of recent studies have reported properties such as small world topology
\cite{Yu08} and critical behavior \cite{Tkacik09,MoraBialek11} exhibited by the inferred
couplings, suggesting interesting physics in the collective dynamics of these biological
networks. One can probably gain more insight into the underlying mechanisms of these
intriguing phenomena by trying to separate the various internal and external components
that contribute to the statistics of the data.

How can we infer connections when neurons are potentially subject to time varying external
input? How does allowing for nonstationarities influence the quality of the model and the
inferred connectivity? Answering these questions quantitatively is the principal aim of
this paper. To do this we fit to neural data a model that does not make any assumptions
about the stationarity of the external field. This is what we call a nonstationary model.
We compare the quality of this model with one that assumes, i.e. forces, stationary
external input to the neurons.  To make this comparison, we evaluate the log likelihood of
the data under the two models.  Since the stationary and nonstationary models have
different number of parameters, we correct the obtained likelihoods to account for this
difference using the Akaike information criterion \cite{Akaike74}.  We also study how the
presence or absence of connections in these models changes the likelihoods.

We perform these analyses on two data sets. One consists of spike trains from 40 neurons
recorded from the salamander retina (courtesy of Michael Berry, Princeton) subject to
stimulation by natural scene movies. The second data set is a set of spike trains from 100
neurons taken from a computational model of a cortical network in a balanced state, also
driven by nonstationary input. More details about these data sets are given in section
\ref{sec:data}.

A model can be optimal in terms of likelihood but fail to capture a specific feature of
the data. We therefore also examine several other statistics evaluated under our models.

Biological neural networks are quite generally dilute: A given neuron is never connected
to more than a small fraction of the other neurons in its local network.  An important
thing to know about such a system is the network graph: just which neurons are connected. 
 We define a noise/signal ratio which measures how well this problem is solved by a given
model.  To calculate it requires knowing the true connections, so we can compute this
statistic only for the model cortical network.

We also study how well the models capture the features seen in two kinds of spike
statistics that have been claimed to be indicators of nontrivial network properties. These
are the frequencies of numbers of synchronous spikes and the frequencies of spike patterns
as a function of their rank (so-called Zipf plots).

For neural data, the probability that $M$ neurons in a population spike simultaneously
typically decays exponentially with $M$.  Independent neurons firing at a fixed rate do
not exhibit this behavior \cite{Schneidman06,Shlens06,Tkacik09}. However, the observed
statistics can be explained by an equilibrium Ising model with time-independent external
field and couplings between neurons \cite{Tkacik09}.

Spike pattern frequencies appear to obey Zipf's law (i.e., the frequency of a pattern is
inversely proportional to its rank).  This has been interpreted, within a stationary model
framework, as a signature of criticality in the network \cite{MoraBialek11}.

Both these statistics may give information about network dynamics, but previous analyses
of them have been based on stationary models.  Here we investigate whether they can be
accounted for as well or better by nonstationary models, with or without couplings between
neurons.

We also consider another problem relevant to the practical use of our models.  Fitting
nonstationary models can be computationally costly. Therefore, we compare different
computational methods for making the fit to the data in the context of the above model
comparisons.  We will compare an exact but potentially slow iterative method for
maximizing the likelihood of the data with two faster mean-field methods.  These have been
tested in a limited way on toy models and artificial data but not previously on biological
data.  We will see that the mean-field methods are quite reliable for the model
comparisons we are interested in here.

The paper is organized as follows. We first describe the data and introduce the kinetic
Ising models used for inferring the connections and the statistics that we use to evaluate
their effectiveness.   We then describe our results and discuss their implications.

\section{Methods}

\subsection{Data}
\label{sec:data}
The analysis reported in the Results section is based on two neural spike train data sets as described below.

\begin{figure}[hbtp]
\centering
\includegraphics[height=3in, width=3in]{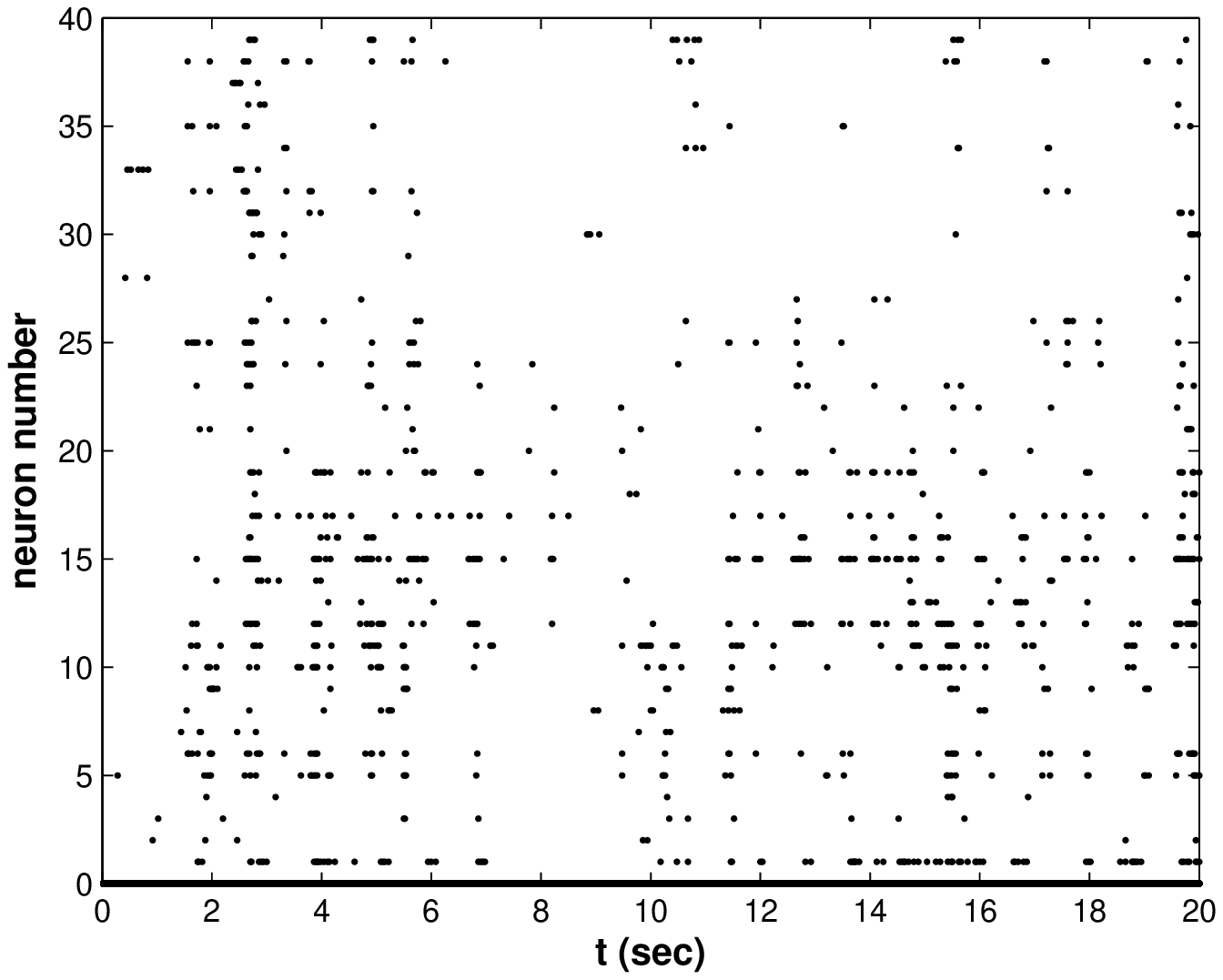}
\includegraphics[height=3in, width=3in]{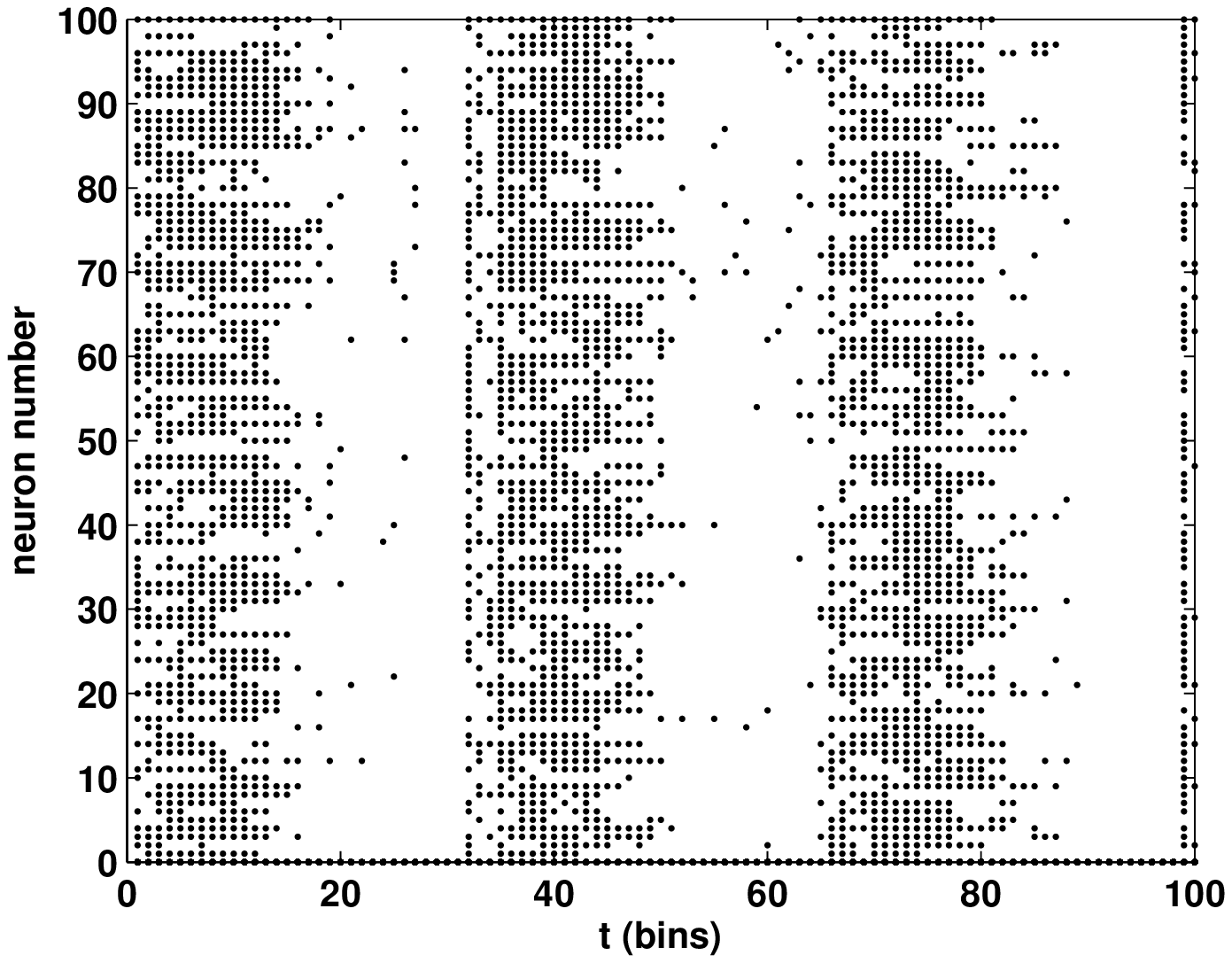}
\caption{Example of spike trains from 40 salamander retinal ganglion cells (left panel) and 100
neurons in the cortical network model (right panel).}
\label{Fig1}
\end{figure}

The first data set, provided by Michael Berry of Princeton University, was recorded from
salamander retina under visual stimulation by a repeated 26.5-second movie clip.  40
neurons were recorded for 3180 seconds (120 repetitions of the movie clip).   Their firing
rates ranged from a minimum of 0.28 Hz to a maximum of 5.42 Hz, with a mean of 1.356 Hz. 
For most of the analysis reported here, the data were binned using 20-ms bins.   We also
used 2-, 5-, and 10-ms bins for some of the calculations, as described later. 
Fig.~\ref{Fig1}a shows spike rasters from a 5-second portion of the data.

We chose 20-ms for the main analysis because these were the smallest ones for which the
estimated time-dependent firing rates (obtained for each bin by averaging the spike counts
in the120 trials) appeared continuous or nearly so.

The second data set was from a fairly realistic computational model of a small cortical
network. The model is described in detail in \cite{Hertz10}  Here we list its main
features.  It contained 800 excitatory and 200 inhibitory neurons, with
Hodgkin-Huxley-like intrinsic conductances and conductance-based synapses.  These 1000
neurons were driven by a further 1600 Poisson-firing neurons which were not part of the
network.  Of the 1600 external neurons, 800 fired a a constant low rate (1 Hz), serving as
a model of the background activity of ``the rest of the brain''.  The rate of the other
800 (representing sensory input) was modulated by a truncated sinusoidal function with a
modulation rate of 3 Hz.  This pulsed input is nonzero slightly more than half (53.3\%) of
the time.  Fig.\ \ref{Fig1}b shows the spike rasters of 100 of the neurons over a 1-second
period. All connections, both those between neurons in the network and from the external
populations to neurons in the network, were randomly diluted, with a 10\% connection
probability.

The strengths of the conductances were chosen so that, when firing, the network was in a
high-conductance state, with effective membrane time constants around a millisecond, as described by Destexhe and coworkers \cite{DestexheNRN03}.  There was no variation in the magnitudes of the synaptic conductances within a class (excitatory-excitatory, excitatory-inhibitory, etc.) beyond that implied by the random dilution, although there was random variation in their temporal characteristics.

A balanced state in this network requires very strong inhibitory synapses to balance the excitatory ones, which are four times more numerous in this network.   The model we will use to fit the data (see below) does not have conductance-based synapses, so to gauge how much stronger the inhibitory synapses are, it is useful to compare effective current-based synaptic strengths, computed from conductances $g$ as $g(V_{rev} - \overline{V})$, where $V_{rev}$ is the reversal potential for the synapse (excitatory or inhibitory) and $\overline{V}$ is a typical value of the membrane potential in the balanced state.  For this network, the ratio of the (absolute value of the) inhibitory effective couplings to the excitatory ones, estimated in this way, is about 4 for excitatory to inhibitory synapses and about 7 for excitatory-to-excitatory ones.   

We collected spike data for 4350 seconds of simulated time and binned the spike trains using 10-ms bins. (Here, our choice of bin size was dictated by our prior knowledge of the range of the synaptic time constants in the model network.) The excitatory neurons had a mean firing rate of 8.77 Hz, and the inhibitory ones had a mean rate of 12.897 Hz.  For the analysis, we used the 100 neurons with the highest firing rates.  Of these, 66 were excitatory and 34 inhibitory.  Their rates ranged from 24.46 Hz to 52.47 Hz, with a mean of 35.43 Hz.  In the analysis, the data were treated as 128 25-second repetitions of a measurement.

For both data sets, having chosen the time bin size, the state of each neuron in each bin is characterized by a binary variable, $S_i(t) = \pm 1$ according to whether neuron $i$  fires or does not fire in time step (bin) $t$. 

\subsection{Kinetic Ising models}

We are interested in inferring a statistical model that maximizes the probability of the spike histories $\{S_i(t)\}_{i=1}^N$, $1 \le t \le T$.  This is different from what one does for Gibbs equilibrium models  \cite{Schneidman06,Roudi09,Roudi09-2}, where one is concerned only with modeling the distribution of spike patterns, irrespective of their temporal order. 

We consider a discrete-time kinetic Ising model.  The state of each neuron is characterized by a binary variable, $S_i(t) = \pm 1$ according to whether neuron $i$  fires or does not fire in time step (bin) $t$.  The dynamics of the model is defined by a simple stochastic update rule: At each time step, neurons receive inputs, $H_i(t)$, from both an external driving field $h_i(t)$ and the other neurons presynaptic to them:
\begin{equation} 
H_i(t) = h_i(t) + \sum_j J_{ij}S_j(t).
\lb{totalinput}
\end{equation}
Each neuron then, independently of all the others, fires at the next time step with a probability, conditional on the current neuron states, which is a logistic sigmoidal function of its total input:
\begin{equation}
{\rm{Pr}}(S_i(t+1)=1|\{s_j(t)\}) = f(H_i(t)),
\lb{DynSK}
\end{equation}
where $f(x) = 1/(1+\e^{-2x})$.  The parameters of this model are the external fields $h_i(t)$ and the couplings $J_{ij}$. For deriving the learning rules for $J_{ij}$ and $h_i(t)$, it will be convenient to write Eq.(\ref{DynSK}) in the form
\begin{equation}
{\rm{Pr}}(S_i(t+1)|\{s_j(t)\}) = \frac{\exp[S_i(t+1)H_i(t)]}{2 \cosh H_i(t)}.
\lb{DynSK2}
\end{equation}

If $h_i(t)$ is time-dependent, the network statistics will be nonstationary.  This gives
us the possibility of describing nonstationary data, provided it is reasonable to assume
that the $J_{ij}$ do not change in time.  Of course, then we need data from many
repetitions of the history of the network to be able to find the $h_i(t)$.
It is important to keep in mind that the $h_i(t)$ represent more than just the external
stimulus controlled by the experimenter. In addition to the external stimulus, they
represent all the input to the recorded neurons from other neurons. It will not in general
be possible to account for the correlations between these neurons in terms of the
spatiotemporal variation of the stimulus, because they may interact with each other.

It will be useful to compare these models with independent-neuron ones, which are simply defined by
(\ref{DynSK}) and (\ref{DynSK2}) with all $J_{ij} = 0$.

It is also possible to formulate an asynchronous-update version of the stationary model 
\cite{Zeng11}.  In this case, if the fields $h_i$ are constant in time and the $\sf J$
matrix is symmetric, the network relaxes to the stationary Gibbs distribution model of
\cite{Schneidman06}.

\subsection{Objective function}

We assume that the data consist of $R$ ``trials'' or repetitions of the network evolution,
each of length $L$ time steps. Accordingly, we denote the state of neuron $i$ at time step
$t$ in trial $r$ by $S_i(t,r)$.  The objective function to be maximized in fitting the
model parameters is the log-likelihood of the data $\{S_i(t,r)\}$ under the model,
\begin{equation}
{\cal L} = \sum_{i=1}^N \sum_{r=1}^R\sum_{t=1}^L \{S_i(t+1,r)H_i(t,r) - \log [2 \cosh H_i(t,r)]\},
\lb{LL}
\end{equation}
where the $H_i(t,r)$ depend on the $S_i(t,r)$ in the same way as in Eq. (\ref{totalinput}) and $N$ is the number of neurons.  

\subsection{Exact algorithm}

We find the $h_i(t)$ and $J_{ij}$ by gradient ascent on Eq. (\ref{LL}):
\begin{eqnarray}
\delta h_i(t) &=&  \frac{\eta}{L} \frac{\partial {\cal L}}{\partial h_i(t)} =  \frac{\eta}{L} \langle S_i(t+1,r) - \tanh(H_i(t,r)) \rangle_r			\nonumber 	\\
&=&  \frac{\eta}{L} [m_i(t+1) - \langle \tanh(H_i(t,r))\rangle_r ]		\lb{exacth}	 \\
\delta J_{ij} &=& \frac{\eta}{L} \frac{\partial {\cal L}}{\partial J_{ij}} = \eta \langle [S_i(t+1,r) -\tanh(H_i(t,r))]S_j(t,r) \rangle_{r,t},
\lb{exactlearning}
\end{eqnarray}
where $m_i(t) = \langle S_i(t,r)\rangle_r$ is the trial mean of $S_i(t,r)$ and $\eta$ is a
learning rate. The averages are over the data -- in (\ref{exacth}) over repetitions for
each time step $t$ and in (\ref{exactlearning}) over both repetitions and time steps.  We
use the term ``exact'' to describe this algorithm in the sense that if the data were
generated by this model, it would recover the parameters exactly in the limit $R \to
\infty$ of infinite data.

The exact algorithm for the stationary model is very similar; the only difference is that
we can regard each time step as a trial.  Then the averages are only over time steps:
\begin{eqnarray}
\delta h_i&=& \frac{\eta}{L} \frac{\partial {\cal L}}{\partial h_i} = \eta \langle S_i(t+1) - \tanh(H_i(t)) \rangle_t			\nonumber 
		\lb{exacthstat}	 \\
\delta J_{ij} &=& \frac{\eta}{L} \frac{\partial {\cal L}}{\partial J_{ij}} = \eta \langle [S_i(t+1) -\tanh(H_i(t))]S_j(t) \rangle_t.
\lb{exactlearningstat}
\end{eqnarray}
It is worth noting that both these algorithms are generally much faster than that for the
stationary Gibbs distribution model.

Choosing the learning rate required a little trial and error.  We found that a learning
rate $\eta =0.05$ was the largest value for which we reliably found monotonic decreases of
our error measures.  Typically, 1000 iterations were necessary to achieve stable errors. 
In the results presented here, all runs were 1000 iterations unless specified otherwise.

For all neurons, there were time bins with no spikes in any trial, and for some neurons
there were bins with spikes in every trial.  Naive inference in this case would lead to
$|h_i(t)| = \infty$.  In these cases, we reduced the empirical $| \langle S_i(t)\rangle |$
by hand from $1$ $0.999$ or, in some cases, to $0.99999$.  These correspond to $|h_i(t)|
\approx 4$ and $6$, respectively.  The results and conclusions we draw from them below do
not appear to be sensitive to this choice.

\subsection{Smoothing}

Our nonstationary models have many parameters (there are $NT$ $h$s), especially when we
use very small time bins, so it can be useful to reduce the effective number of parameters
by smoothing the inferred $h$s in time.  We can do this by subtracting a penalty term
\begin{equation}
{\cal K} = \half \kappa \sum_{it} [h_i(t+1) -h_i(t)]^2
\end{equation}
from $\cal L$.  This leads to an extra term in $\delta h_i(t)$ proportional to $\kappa [h_i(t-1) -2h_i(t) + h_i(t+1)]$.

\subsection{Mean field theories}

Mean field theories provide faster algorithms for inferring network parameters than the
exact learning rules described above.  However, they are approximations.  Therefore in
this paper we apply both exact and mean-field algorithms and compare the resulting
inferred couplings.

We employ two kinds of mean field theories.  We call the simpler of these naive mean field
theory \cite{Roudi11}.  One starts with the learning rule (\ref{exactlearning}) for
$J_{ij}$ at $\delta J_{ij} = 0$ (i.e., after learning is finished), again writing
$S_i(t,r) = m_i(t)+\delta S_i(t,r)$, and expanding the tanh to first order in $\delta S$. 
Then the naive mean field equations
\begin{equation}
m_i(t+1) = \tanh[h_i(t) + \sum_j J_{ij} m_j(t)],
\lb{MFeqns}
\end{equation}
permit elimination of the zeroth order term, and one is left with a set of linear matrix equations:
\begin{equation}
\langle D_{ij}(t) \rangle_t = \sum_k J_{ik} B_{kj}^{(i)},
\lb{MFeqn}
\end{equation}
where
\begin{equation}
B_{jk}^{(i)} = \langle(1-m_i^2(t+1))C_{jk}(t)\rangle_t .
\lb{Bdef}
\end{equation}
Thus, for each neuron $i$, we obtain
\begin{equation}
J_{ij} = \sum_k  \langle D_{ik}(t) \rangle_t [({\sf B}^{(i)})^{-1}]_{kj}.
\lb{MFJ}
\end{equation}

Once the couplings are found in this way, equations (\ref{MFeqns}) can then be solved for the $h_i(t)$, knowing the $m_i(t)$.

Naive mean field theory (henceforth abbreviated nMF) is exact in the limit of weak coupling strength and also for arbitrary coupling in a large, densely-connected system if the mean of the couplings is positive and their standard deviation is not large relative to the mean.   In the opposite limit (standard deviation much larger than mean), fluctuations around the mean field become important, and there is no general simple solution.  One strategy is to expand around the weak coupling limit \cite{Plefka82}, as was done by two of us \cite{Roudi11,Roudi11-2}.  Here we take another route:  There exists an exact solution for densely connected systems if the coupling matrix $\sf J$ is strongly asymmetric ($J_{ij}$ and $J_{ji}$ independently distributed) \cite{Mezard11,Sakellarious12}.  This is a reasonable assumption for randomly-wired neuronal networks, since $J_{ij}$ and $J_{ji}$ represent different synapses.   Following M{\'e}zard and Sakellariou, we abbreviate this full mean-field theory simply as MF.

In this case the internal fields acting on different units are independent Gaussian variables, and (\ref{MFeqns}) are replaced by
\begin{equation}
m_i(t+1) = \int Dx \tanh(b_i(t) + x\sqrt{\Delta_i(t)}) ,				\lb{MMF}
\end{equation}
where
\begin{equation}
\int Dx ( \cdots ) \equiv \int \frac{dx}{\sqrt{2\pi}} {\rm e}^{-\half x^2} ( \cdots )		\lb{defGauss}
\end{equation}
means integrate over a univariate Gaussian, 
\begin{equation}
b_i(t) = h_i(t) + \sum_j J_{ij} m_j(t)									\lb{defb}
\end{equation}
is the internal field from naive mean-field theory, and the internal field variance is
\begin{equation}
\Delta_i(t) = \sum_j J_{ij}^2 (1- m_j^2(t)) .						\lb{varhMMF}
\end{equation}

If we again write  $S_i(t,r) = m_i(t)+\delta S_i(t,r)$ and expand the tanh to first order
in $\delta S$, this time using (\ref{MMF}) instead of (\ref{MFeqns}), we are again led to
an equation of the form (\ref{MFJ}), but now with
\begin{equation}
B_{jk}^{(i)} = \left \langle \int Dx [1- \tanh^2(b_i(t) + x\sqrt{\Delta_i(t)})] C_{jk}(t)\right \rangle_t .
\lb{BdefMMF}
\end{equation}

This calculation has to be done iteratively.  We start with an initial guess for the $J$s
from nMF.  From it we estimate the field variances $\Delta_i(t)$ using (\ref{varhMMF}),
and using these we solve (\ref{MMF}) for the $b_i(t)$ by numerical iteration.  This
enables us to calculate the $B_{kl}^{(i)}$ from (\ref{BdefMMF}) and, from them, the $J$s
using (\ref{MFJ}).  We then use these $J$s to get a better estimate of the $\Delta_i(t)$
and repeat the calculations, leading to new $J$s, and iterate this procedure until the
couplings converge. In practice this took 5-10 iterations of both the outer (successive
estimates of the $J$s) and inner (successive estimates of the $b_i(t)$) iteration loops.

\subsection{Error and quality-of-fit measures}

There are a number of measures of the error or quality of fit.  One is the objective
function itself. We use it, in conjunction with an Akaike penalty, in model comparison. 
Akaike \cite{Akaike74} showed that, under rather general conditions, the log-likelihood
evaluated on the data is a biased (over)estimate of the true log-likelihood of a model
and, furthermore, that this bias is just equal to the number of parameters $k$ in the
model.  Thus, the Akaike penalty we subtract from the empirical log-likelihood in doing
model comparison is simply the number of parameters.

The use of a different penalty term has been proposed in the statistical literature by
Schwarz \cite{Schwarz78}.  He takes a Bayesian approach, according to which the likelihood
of a model is proportional to the integral over its posterior distribution.  Assuming a
flat prior and expanding the log-likelihood around its maximum, the quadratic term is
proportional to the sample size $n$ and the integral to be done is over a $k$-dimensional
Gaussian.  The result is proportional to $n^{k/2}$. Taking the log of this gives Schwarz's
penalty, $k \log \sqrt{n}$.  For large sample size, it penalizes large models more
strongly than Akaike's does.

The merits of these criteria, commonly called AIC (Akaike information criterion) and BIC
(Bayesian information criterion), and others related to them are still discussed in the
statistical literature \cite{BurnhamAnderson02}.  Here we generally use Akaike's, but in
one case we also employ Schwarz's and compare the conclusions they lead to.

The objective function is extensive in the number of neurons and the length of the data
set.  When we report computed values here (with or without Akaike or Schwarz penalties),
they are  per neuron per time step.  We use natural logarithms; dividing our numbers by
$\log 2$ expresses them in bits.

For the data generated by the computational cortical network, we know which connections
are actually present, so we can also evaluate the following measure of the accuracy with
which the true connections are found by the model.  Consider the inhibitory synapses.  In
the computational network as implemented here, the strengths of their conductances are
always the same if they are nonzero.  The differences in their temporal characteristics
are beyond the scope of our simple memory-less Ising model, so we have to disregard them
here.  We would then hope that the algorithm would find about the same (negative) value
for $J_{ij}$ when there is an inhibitory synapse from $j$ to $i$ and zero otherwise. 
However, because of finite data and model mismatch, the $J$s found which should be
negative are actually spread around some mean value $-J_0$, with standard deviation
$\sigma_1$, and those which should be zero are actually spread around zero with some
standard deviation $\sigma_0$.   We therefore adopt as a measure of the network
reconstruction error the noise-signal ratio
\begin{equation}
\zeta = \frac{\sigma_1 + \sigma_0}{J_0}.
\lb{nsr}
\end{equation}
\subsection{Other statistics}

For both data sets and both kinds of models, with and without couplings, we also consider
two more simple statistics.  The first is the estimated probability of different numbers
of synchronous (i.e., within the same time bin) spikes.  Defining
\begin{equation}
M(t,r) = \half \sum_i [1+S_i(t,r)],
\end{equation}
the estimated probability of $M$ synchronous spikes is 
\begin{equation}
P(M) = \frac{1}{RT}\sum_{r,t} \delta_{M(t,r), M}.				\label{PofM}
\end{equation}

To construct the other statistic, we compile a list of the number of times every observed
spike pattern $S_i(t,r)$ occurs in the data.  We then rank-order this list by the sizes of
these counts and plot the counts against the rank (a so-called Zipf plot).

\section{Results}

\subsection{Model comparison}

\subsubsection*{Salamander retinal data}

\begin{figure}[hbtp]
\centering
\subfigure{\includegraphics[height=3in, width=3in]{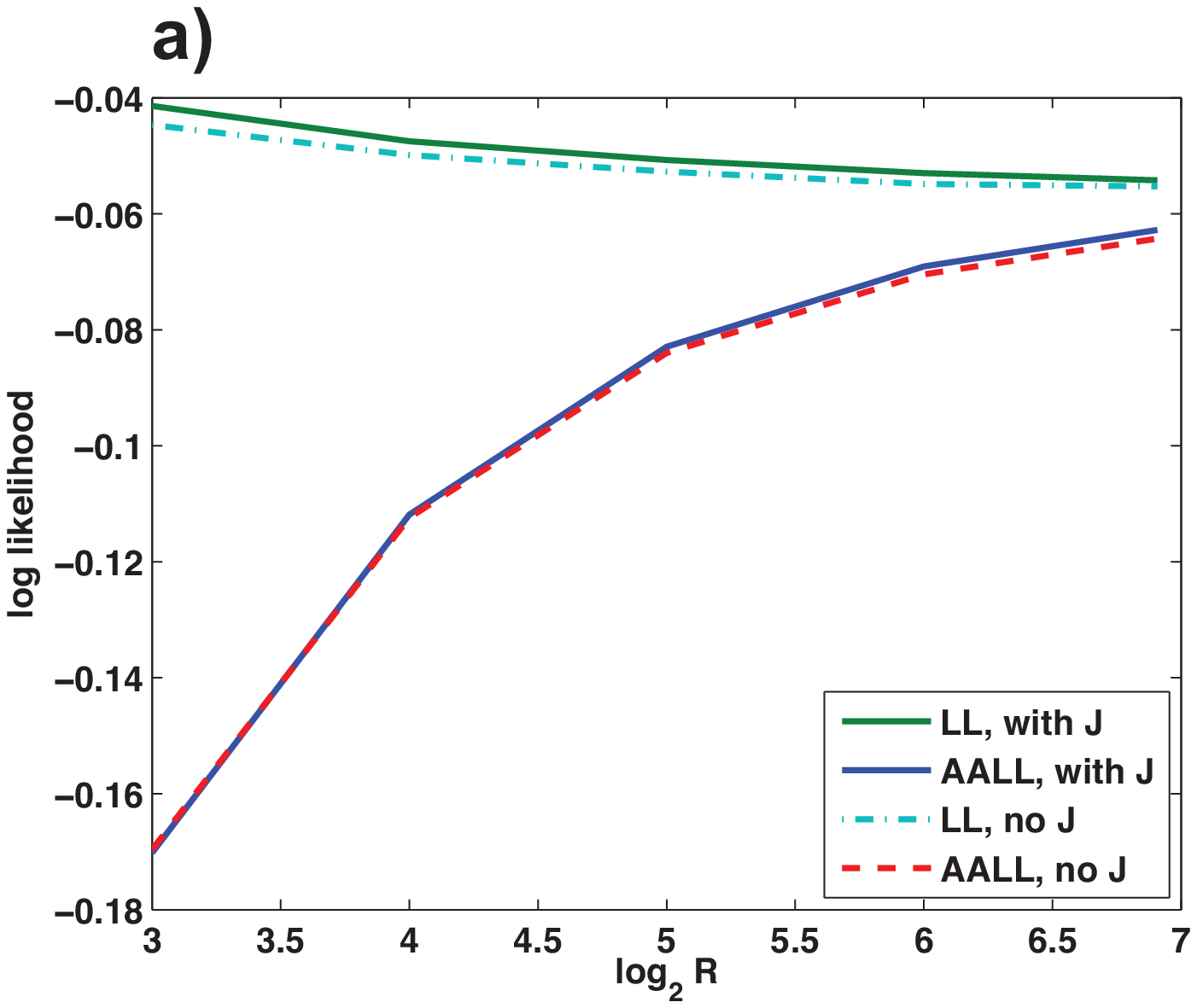}}
\subfigure{\includegraphics[height=3in, width=3in]{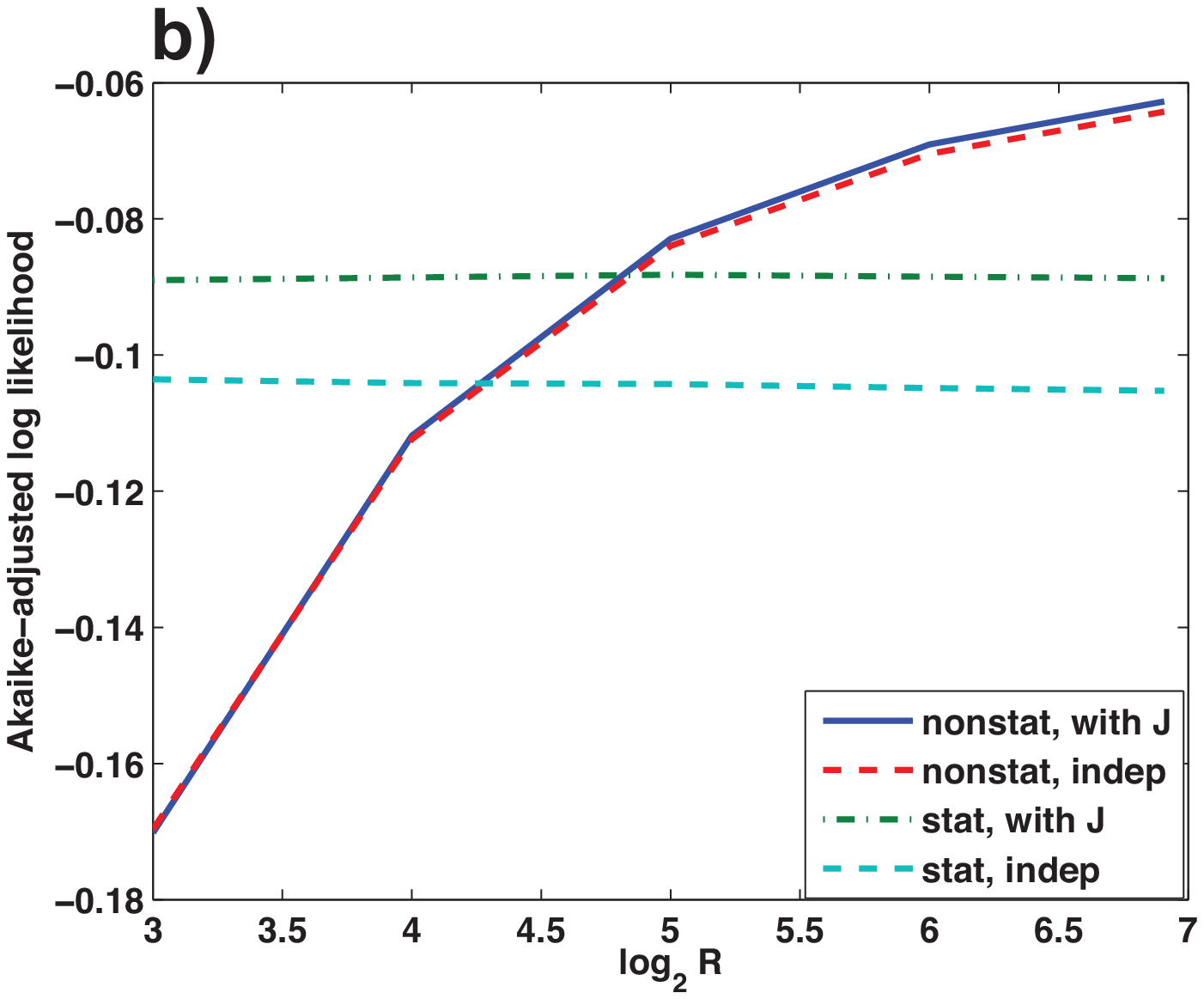}}
\subfigure{\includegraphics[height=3in, width=3in]{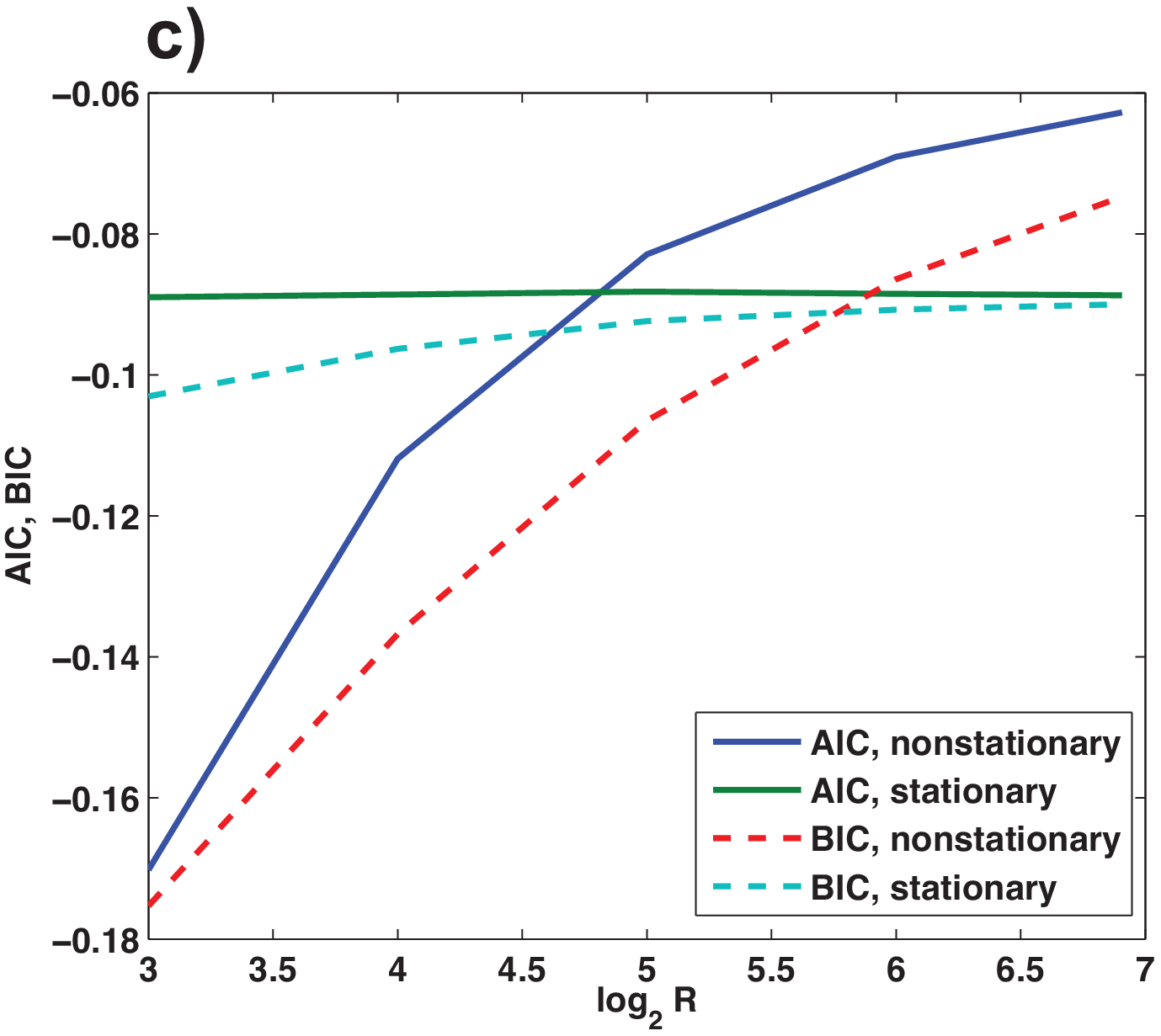}}
\caption{(a) Log likelihoods (per neuron, per bin) of retinal ganglion cell spike trains
under nonstationary models with and without couplings, as functions of number of
repetitions $R$ of the 26.5-s movie clip.  The upper pair of curves are the raw log
likelihoods, and the lower pair are the Akaike-corrected values.  Within each pair, the
higher curve is for the model with couplings and the lower is for an independent-neuron
model.   (b) Akaike-corrected log likelihoods under stationary and nonstationary models,
with and without couplings.  (c) Comparison of model comparisons (with couplings): AIC
(solid lines) vs BIC (dashed lines).}
\label{Fig2}
\end{figure}

We analyzed the fit of the kinetic Ising model to the data set of 40 ganglion cells in a
salamander retina. We calculated the log likelihoods for this data in the exact
nonstationary algorithm for different data sets sizes. Fig.\ \ref{Fig2}a shows log
likelihoods, with and without Akaike penalties, for nonstationary models with and without
$J$s.  For these data, the model with couplings is only slightly better than the
independent neuron model: Almost all the variation in the data could be accounted for by
the time-dependent inferred fields $h_i(t)$.  Although for small $R$ taking into account
Akaike adjustment has a big influence on the value of the log-likelihood, it
loses its importance when enough repeats are present.

This was not true for stationary models, as can be seen in Fig.\ \ref{Fig2}b.  Stationary
models with $J$s are clearly always favored over those without, and for limited data
(fewer than 28 repetitions) they are also favored over the nonstationary models.

Comparing the stationary and nonstationary models, the log-likelihood of the data under
the stationary model is significantly less than the one under nonstationary model. One
might argue that this is due to the fact that the nonstationary model without couplings
has a lot more parameters than the stationary model with couplings. In fact this argument
is correct when only a small number of repeats are used for inferring the nonstationary
input: for small $R<28$ the nonstationary model performs worse than the stationary one
when the Akaike penalty is taken into account.  However, when there are enough repeats
($R>28$), the situation reverses: the nonstationary models, with or without couplings,
outperform the stationary model with couplings even after Akaike corrections.

As mentioned above, Schwarz's Bayesian approach penalizes models with many parameters
(such as our nonstationary ones) more severely than Akaike's.  We therefore also applied
the Schwarz penalty to the models with couplings.  Fig.\ \ref{Fig2}c shows both the Akaike
and Schwarz stories:  Under the Bayesian criterion, one needs data from at least 55
stimulus repetitions to conclude that the nonstationary model is superior, while under
Akaike's criterion only about half as many repetitions were necessary.  However, the
conclusion  based on the entire data set available here (120 repetitions) is the same: 
the nonstationary model fits the data better.

\begin{figure}[hbtp]
\centering
\includegraphics[height=3in, width=4in]{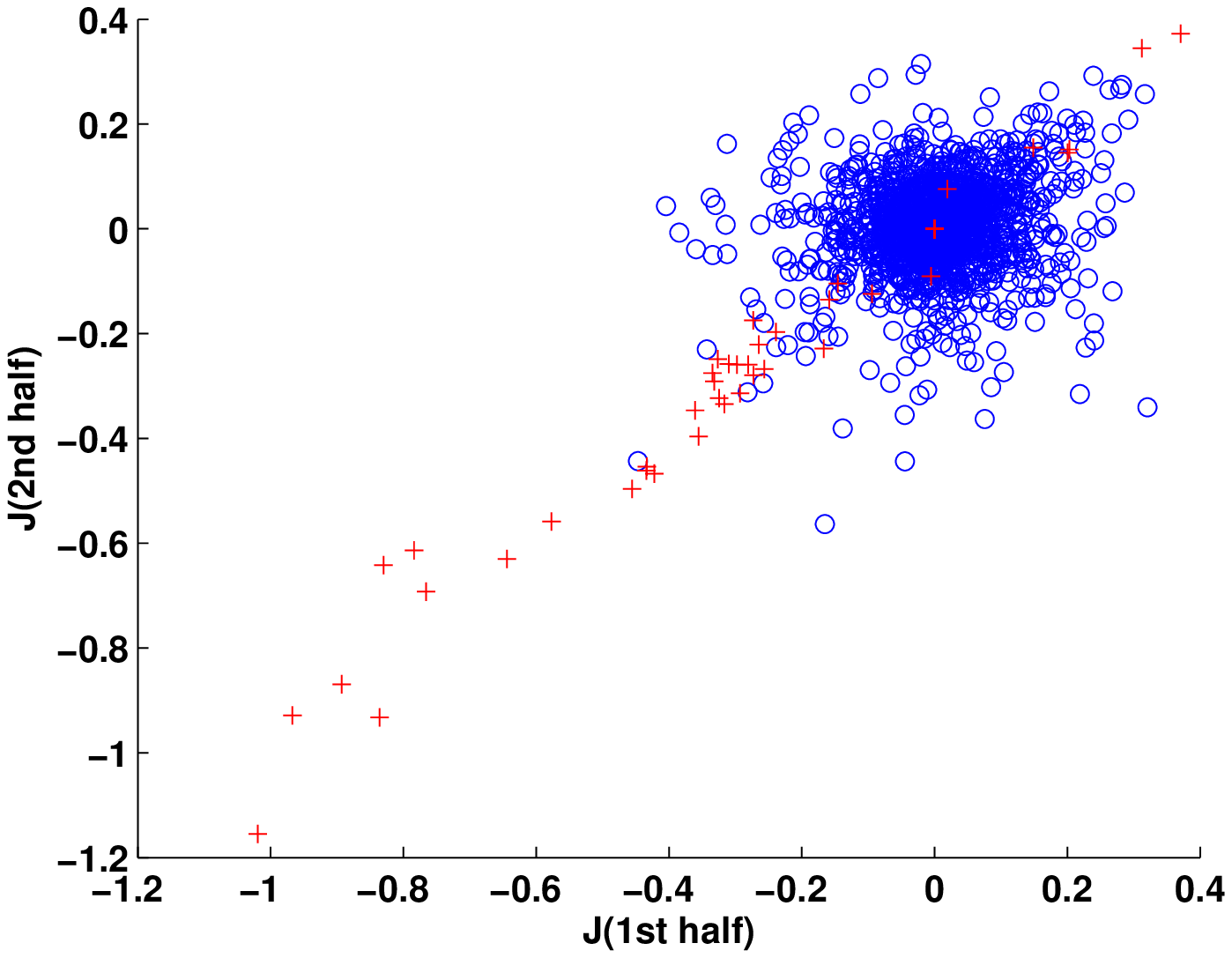}
\caption{ Couplings $J_{ij}$ inferred from salamander retinal data using a nonstationary
model: Couplings inferrred from the first 60 repetitions of the stimulus movie clip
plotted against those inferred from the second 60 repetitions.  Blue circles indicate
couplings between neurons ($J_{ij}$, $i \neq j$, and red crosses indicate self-couplings
($J_{ii}$).}
\label{Fig3}
\end{figure}

The evident insignificance of the $J$s in the nonstationary model is also clarified by
Fig.\ \ref{Fig3}.  Here we performed the inference (using the exact algorithm) separately
for the first and second halves of the data and scattered the results $J$s against each
other.  Fig.\ \ref{Fig3} shows that almost all significant $J$s are inhibitory
self-couplings, reflecting an apparent refractory tendency of the neurons.  Couplings
between neurons are small, and no systematic relation between those found from the two
halves of the data is apparent.

We also inferred nonstationary models for the entire data set using smaller time bins: 2
and 10 ms.  Since the number of $h$s in the model is inversely proportional to the bin
size, we used the smoothing technique described in Sect.\ 2.5, adjusting the smoothing
parameter $\kappa$ to keep the effective number of parameters approximately the same as
for the 20-ms calculations without smoothing.   Thus, the log likelihoods of these models
could be compared directly with the 20-ms models and with each other (their Akaike
penalties were approximately equal).  We found that all of the models using smaller bins
were inferior to the 20-ms model:  The log likelihoods (per 20 ms) were $-0.124$,
$-0.073$, and $-0.051$ for 2-, 10- and 20-ms bins, respectively. Increasing the 
size of the time bins, on the other hand, increases the likelihood. Eventually the 
log-likelihood converges to zero when the bin size encompasses the whole data set 
and the single bin is occupied. The model, however, is then a trivial one.

\begin{figure}[hbtp]
\centering
\subfigure{\includegraphics[height=2in, width=2in]{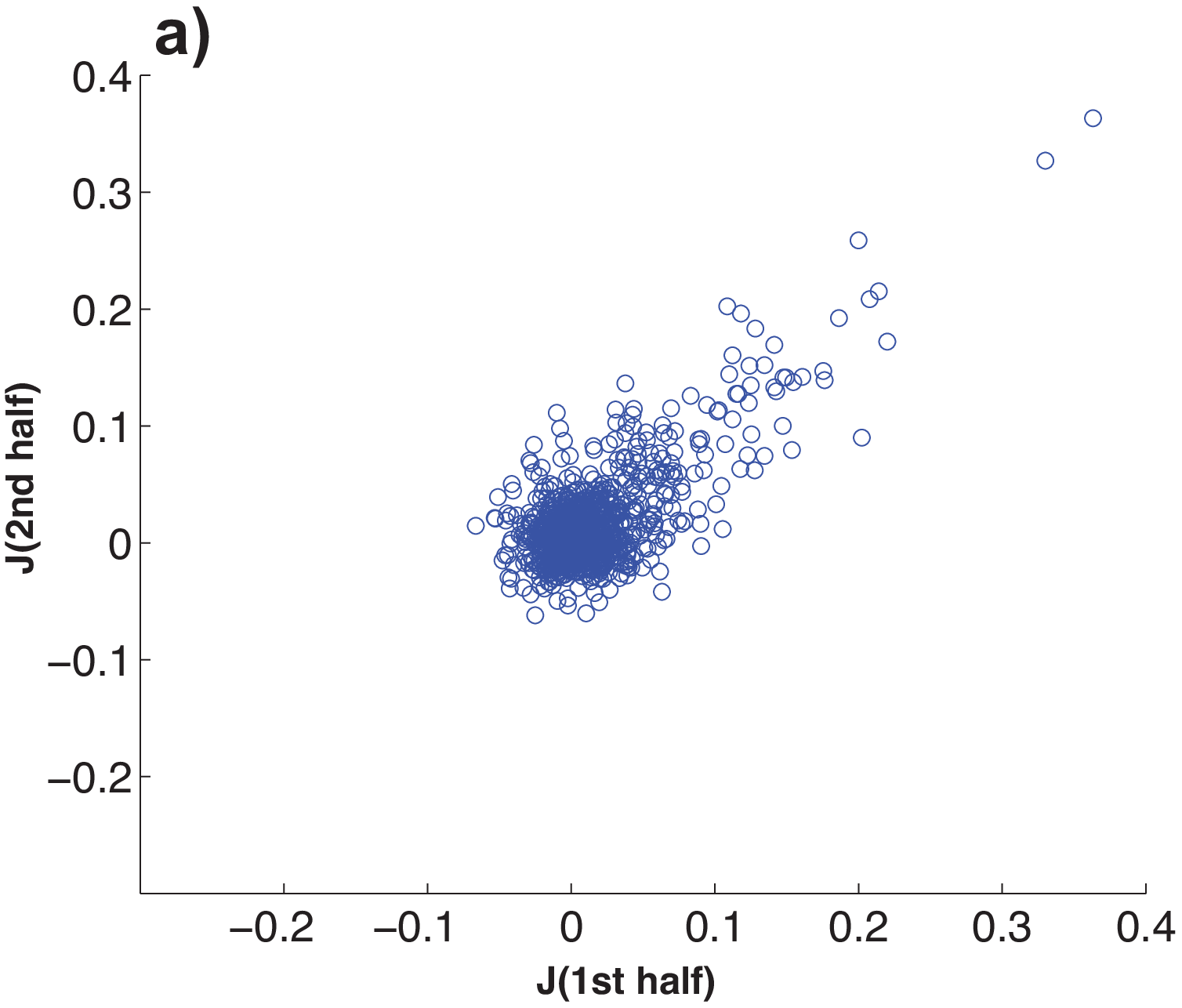}}
\subfigure{\includegraphics[height=2in, width=2in]{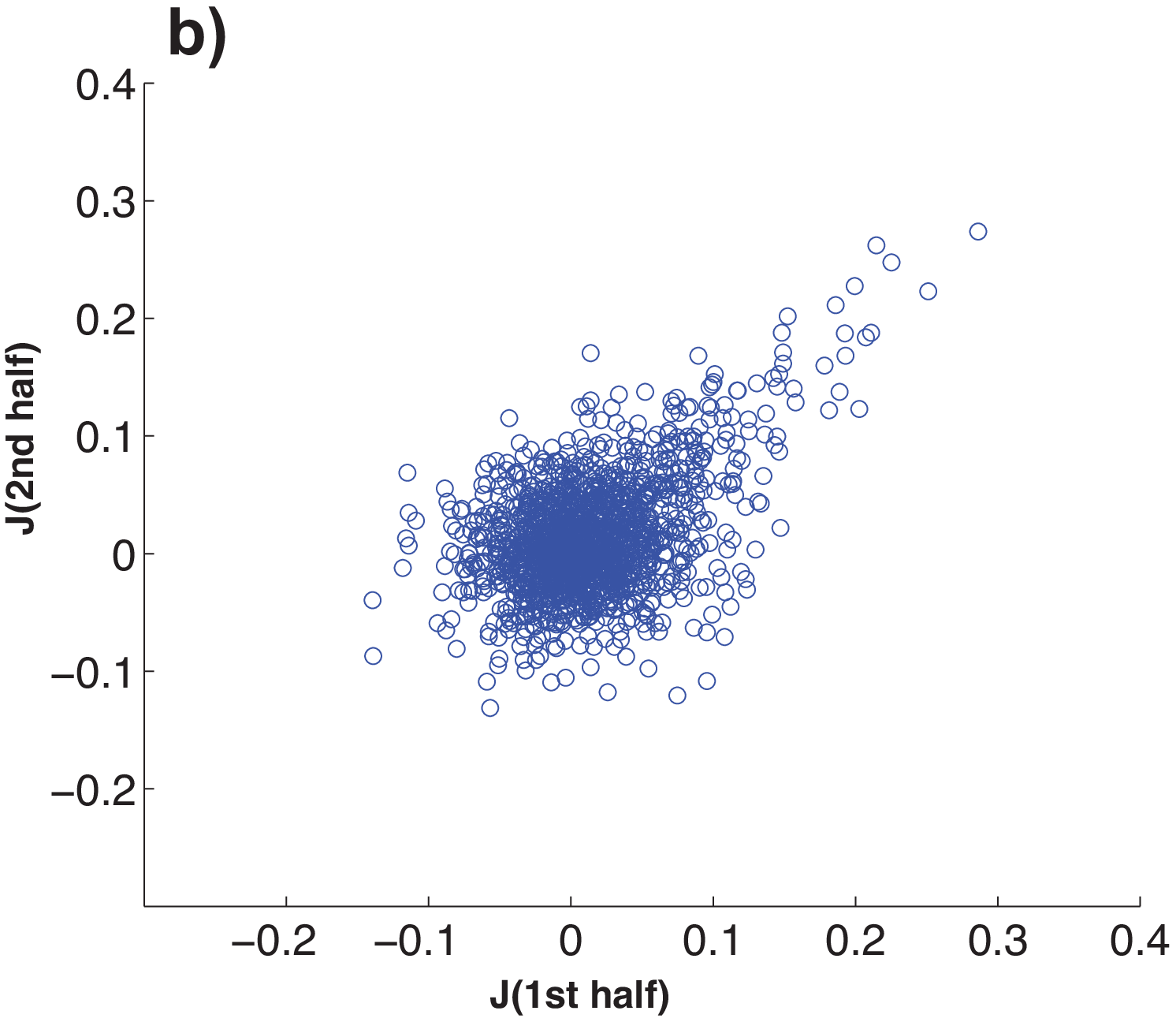}}\\
\subfigure{\includegraphics[height=2in, width=2in]{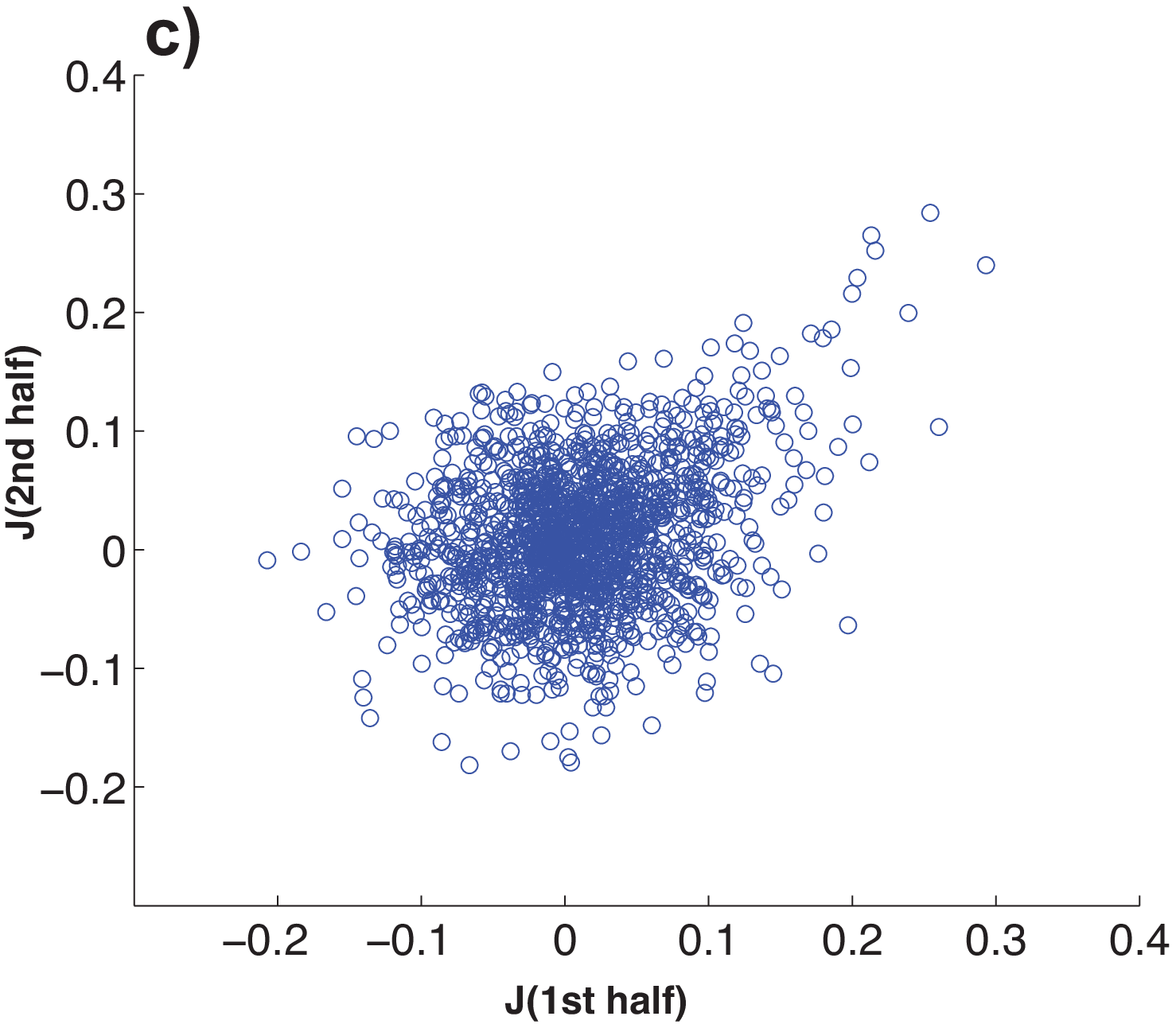}}
\subfigure{\includegraphics[height=2in, width=2in]{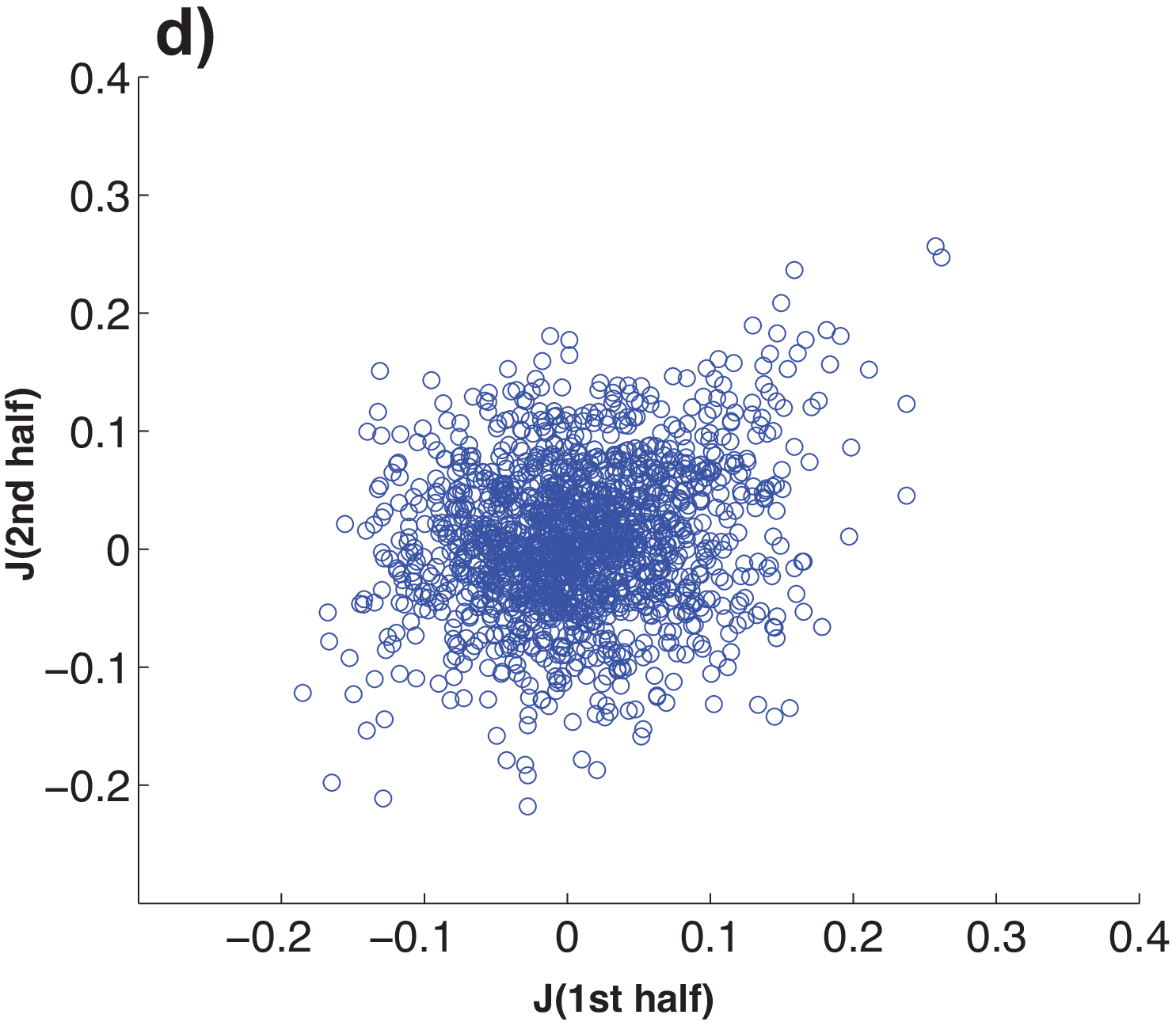}}
\caption{Off-diagonal couplings $J_{ij}$ inferred from the first 60 repetitions of the
stimulus movie clip plotted against those inferred from the second 60 repetitions for
salamander retinal data using nonstationary models based on (a) 2, (b) 5, (c) 10 and (d)
20 ms.}
\label{Fig3X}
\end{figure}

Nevertheless, the smaller-bin models revealed something interesting when we compared the
$J$s inferred from the two halves of the data, as shown in the graphs in Fig.\
\ref{Fig3X}.  They show that some credible $J$s, all positive, are inferred for the
smaller time bins (2 and 5 ms).   Perhaps these couplings are also present for larger time
bins, but there they are lost inw the noise of the many spurious inferred $J$s. Comparison
of the inferred $\sf J$ matrices and their transposes (not shown) revealed that these
couplings are largely bidirectional. We note, however that, the presence of
statistically significant Js for small bin sizes might also be a consequence of the
regularization we imposed on the variation of the fields in order to limit model
complexity. Indeed, for small bin sizes, the regularizer suppresses correlated fluctuations 
of fields at high frequency. High frequency correlated fluctuations in the spins can
therefore be explained, within the regularized model, only by non-zero couplings.

\begin{figure}[hbtp]
\centering
\subfigure{\includegraphics[height=1.5in, width=1.8in]{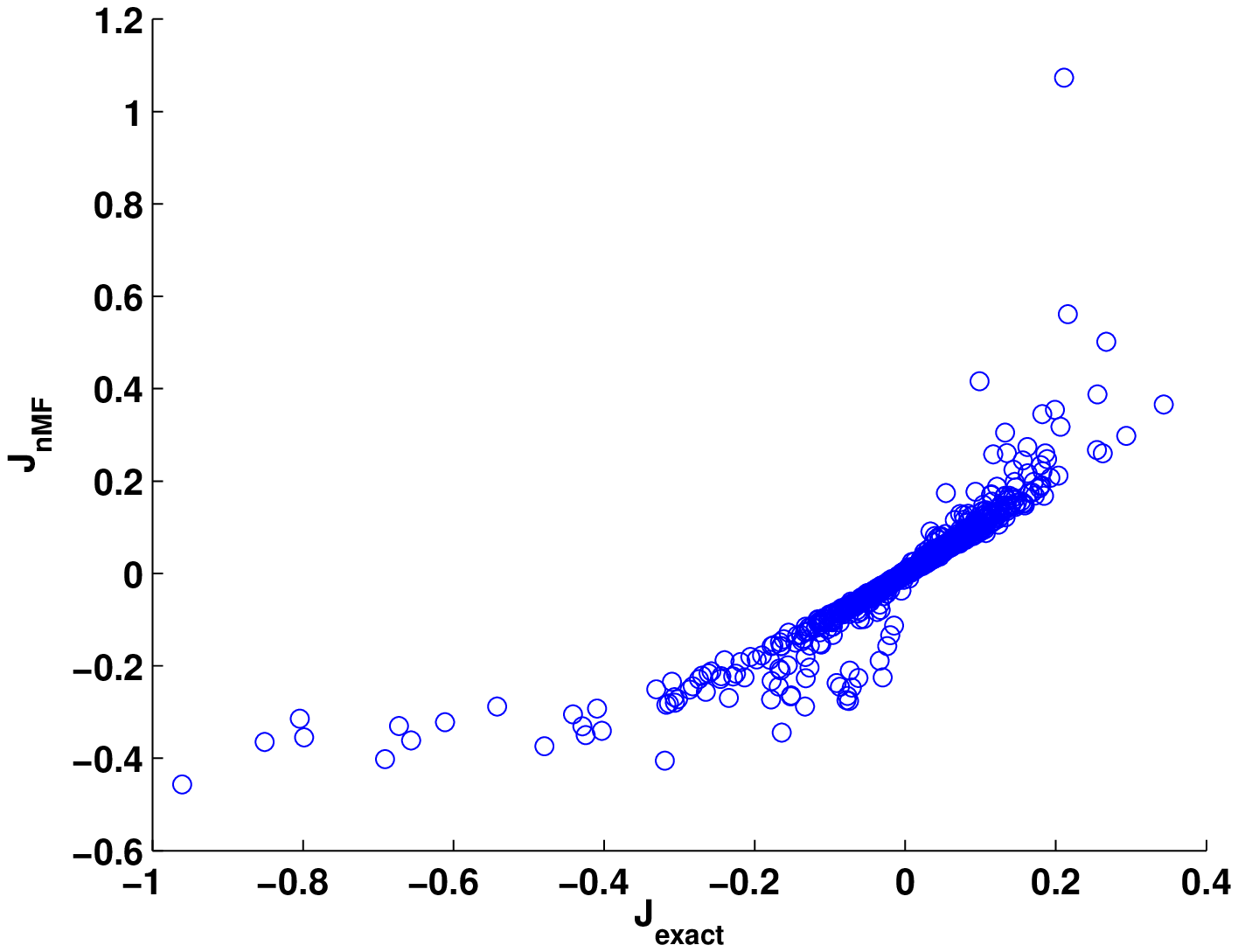}}
\subfigure{\includegraphics[height=1.5in, width=1.8in]{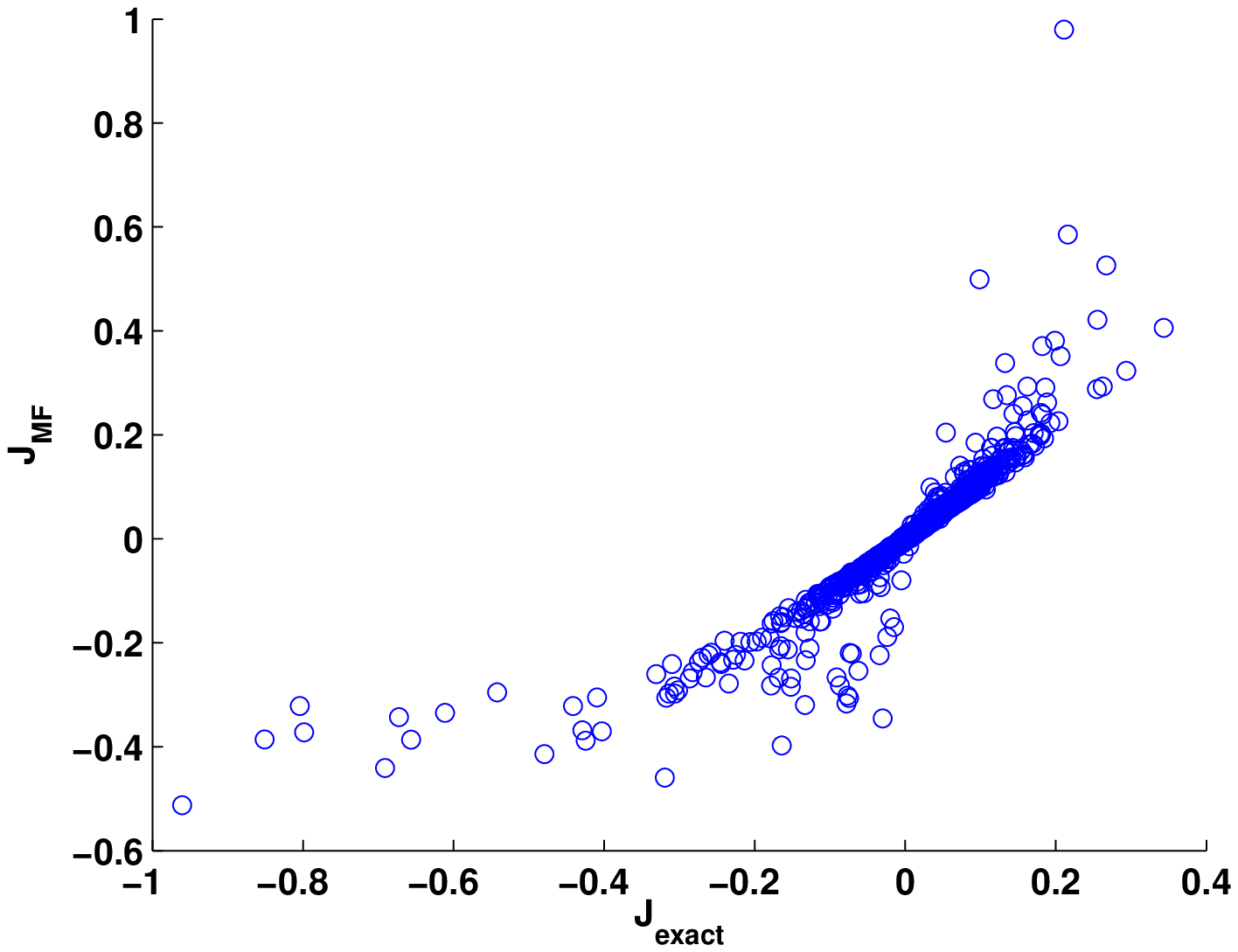}}
\subfigure{\includegraphics[height=1.5in, width=1.8in]{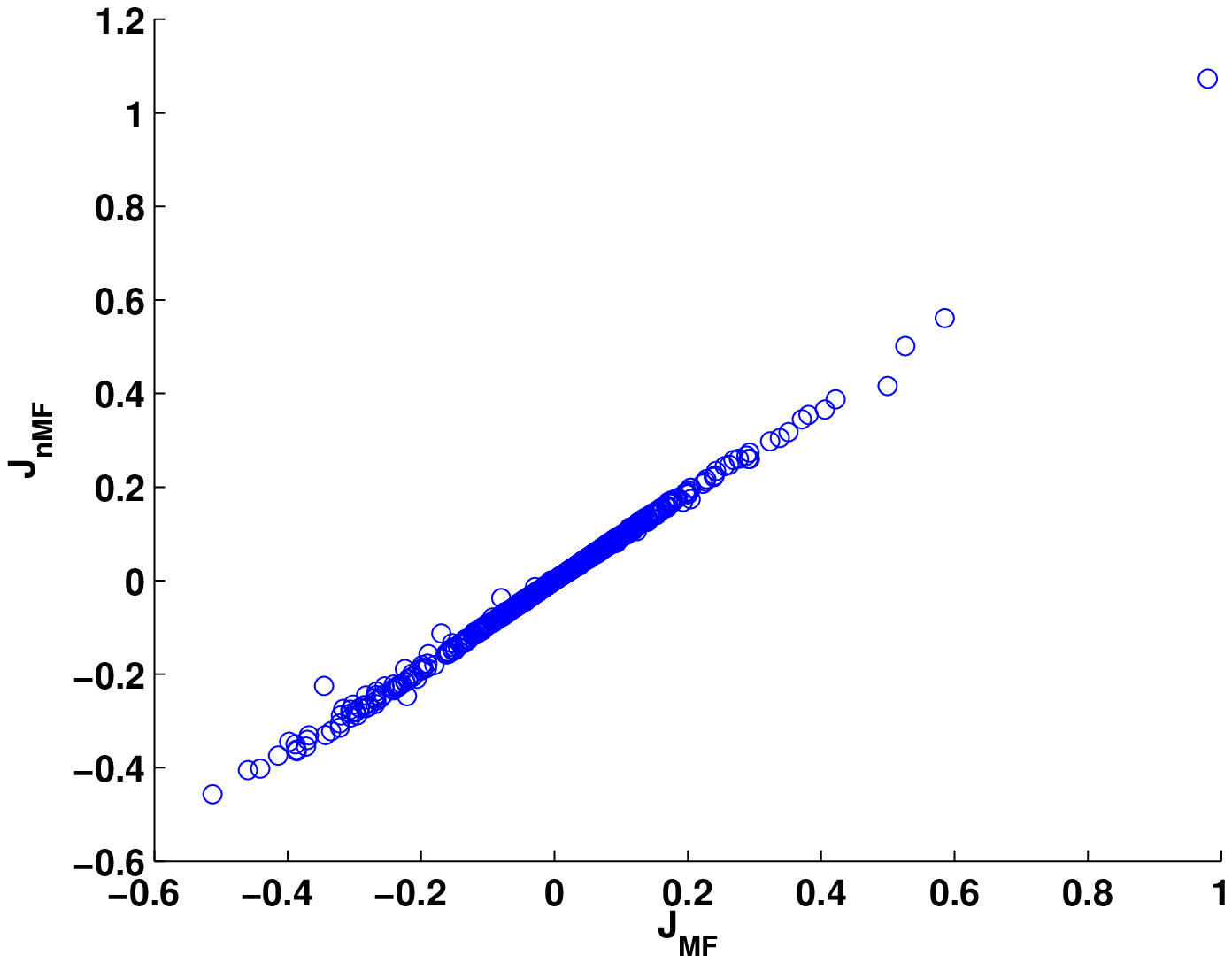}}
\caption{Comparison of couplings in nonstationary model inferred from retinal data by
exact  and mean field algorithms: exact algorithm vs nMF, exact algorithm vs MF, and nMF
vs MF.}
\label{Fig4}
\end{figure}

Returning to the 20-ms models, we also compared exact and mean-field algorithms on these
data. One can see in Fig.\ \ref{Fig4}a-c that nMF and MF agree qualitatively with the
exact algorithm, except that they systematically overestimate large positive (and, to a
lesser extent, large negative) $J$s.   These differences appear to make very little
difference in the estimates of the log likelihoods.  For the exact algorithm, nMF, and MF,
respectively, the log likelihoods (with the Akaike penalty) of the full nonstationary
model on the complete data (120 repetitions) were -0.062748, -0.062872, and -0.062823. 
The differences are at most 0.2\% or less, so all our conclusions above about model
comparison can be drawn equally well from very fast (at most a few minutes) mean-field
calculations as from the lengthy (several hours) calculations using the exact algorithm.

\subsubsection*{Model cortical network}

\begin{figure}[hbtp]
\centering
\subfigure{\includegraphics[height=3in, width=3in]{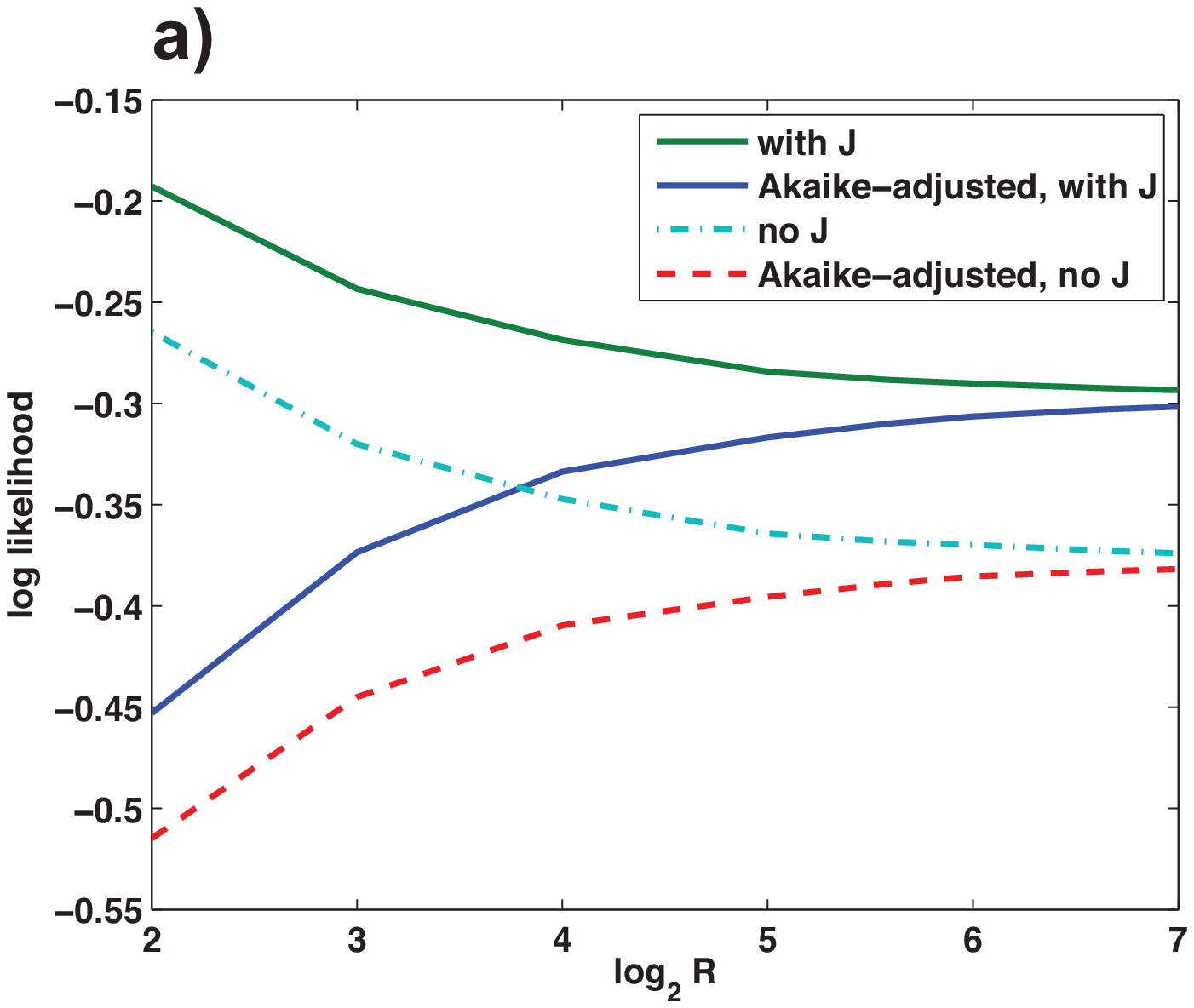}}
\subfigure{\includegraphics[height=3in, width=3in]{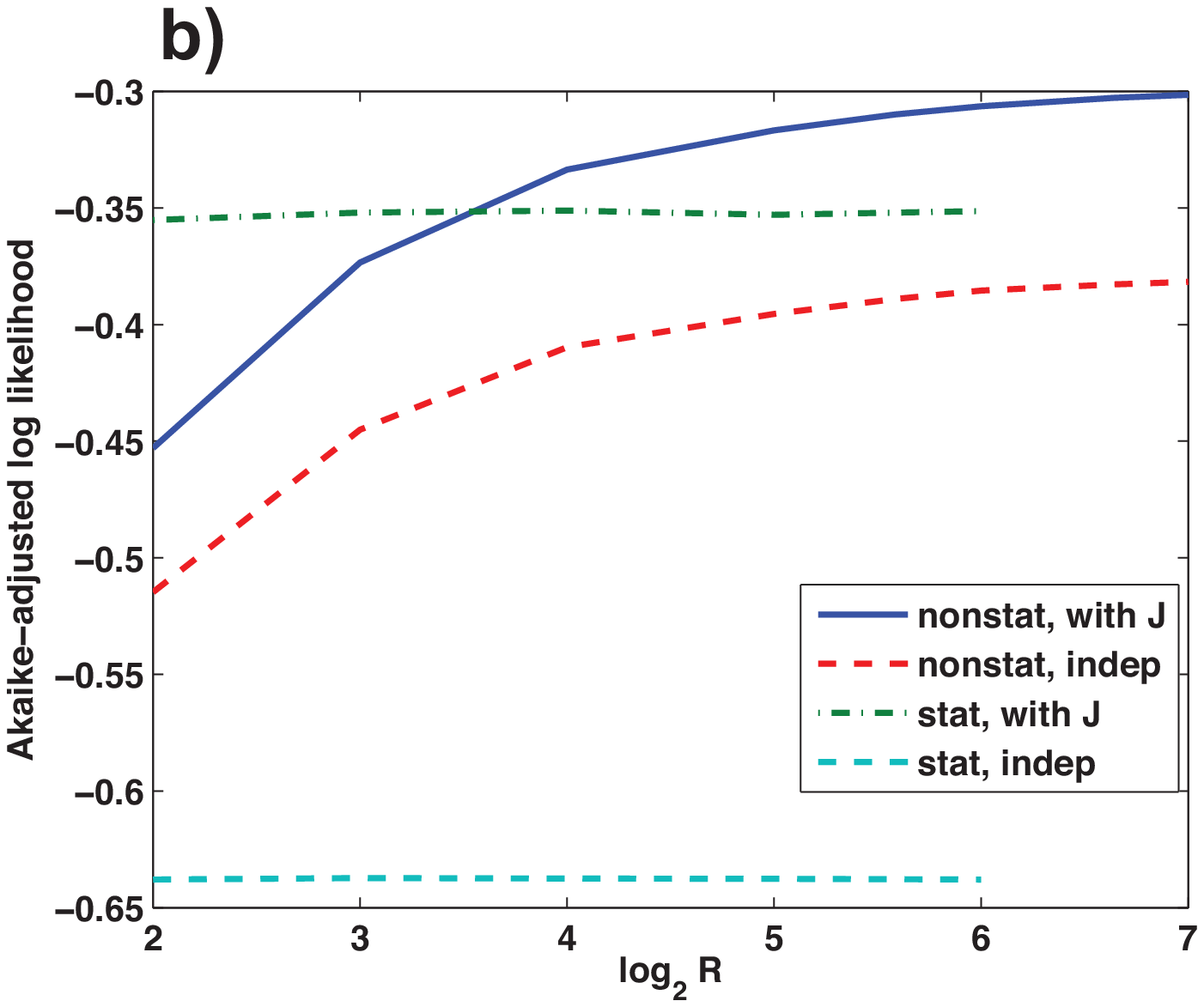}}
\caption{ (a) Log likelihoods of the cortical model data as a function of number of
stimulus repetitions under nonstationary models with and without couplings and with and
without Akaike corrections.  (b) Akaike-corrected log likelihoods, as functions of number
of stimulus repetitions, under stationary and nonstationary models with and without
couplings. }
\label{Fig5}
\end{figure}

We also investigated the fitting of our kinetic Ising model to data generated by our small cortical
network model for which we could generate as much data as we wanted
and for which the true connections and external field was known to us. Similar to the retinal data,
as a quality-of-fit measure, we use the log-likelihood of the data under the kinetic
Ising model. We do the analysis using the exact nonstationary algorithm for data sets of 8 up to 128
repetitions. Fig.\ \ref{Fig5}a shows the log-likelihoods with and without the Akaike penalty as a function of
the number of repetitions. The log-likelihoods are calculated with and without (independent-neuron
model) the couplings $J_{ij}$. It is evident that the model with the $J$s is better than the
one without them. In both cases, as the number of repetitions increases, the Akaike correction
becomes less important. The same Akaike-adjusted log-likelihoods, with and without $J$s, are shown
in Fig.\ \ref{Fig5}b, together with the corresponding results for a stationary model. It is evident that the
nonstationary independent model is much better than the stationary independent one for all numbers
of repetitions. The quality of the nonstationary model with $J$s in comparison to the stationary one
with $J$s depends on the number of repetitions. The nonstationary one has higher Akaike-adjusted
log-likelihood when the number of repetitions is greater than 11. This shows that the size of the
data set can be significant in the choice between models.

\begin{figure}[hbtp]
\centering
\subfigure{\includegraphics[height=1.4in, width=1.8in]{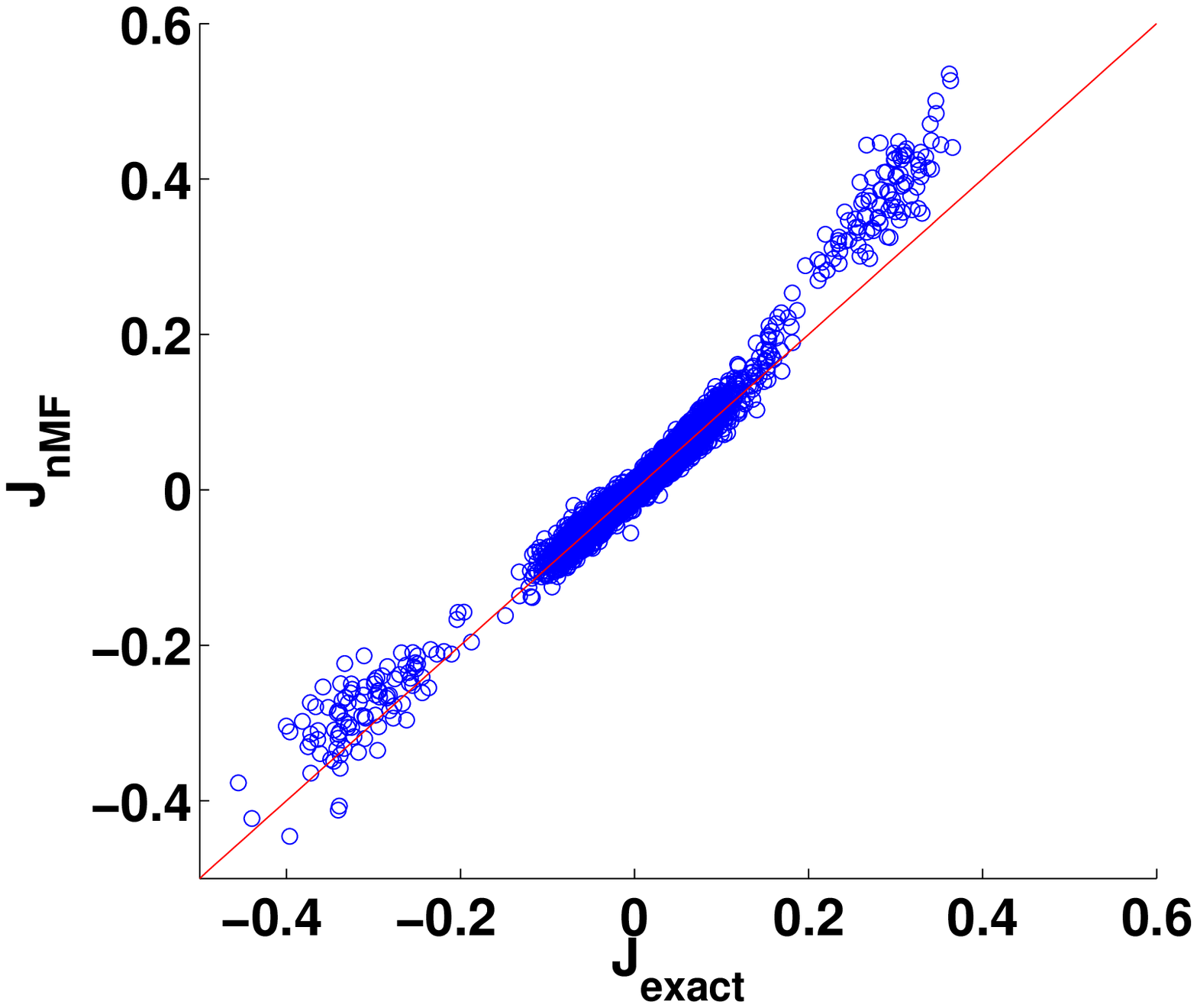}}
\subfigure{\includegraphics[height=1.4in, width=1.8in]{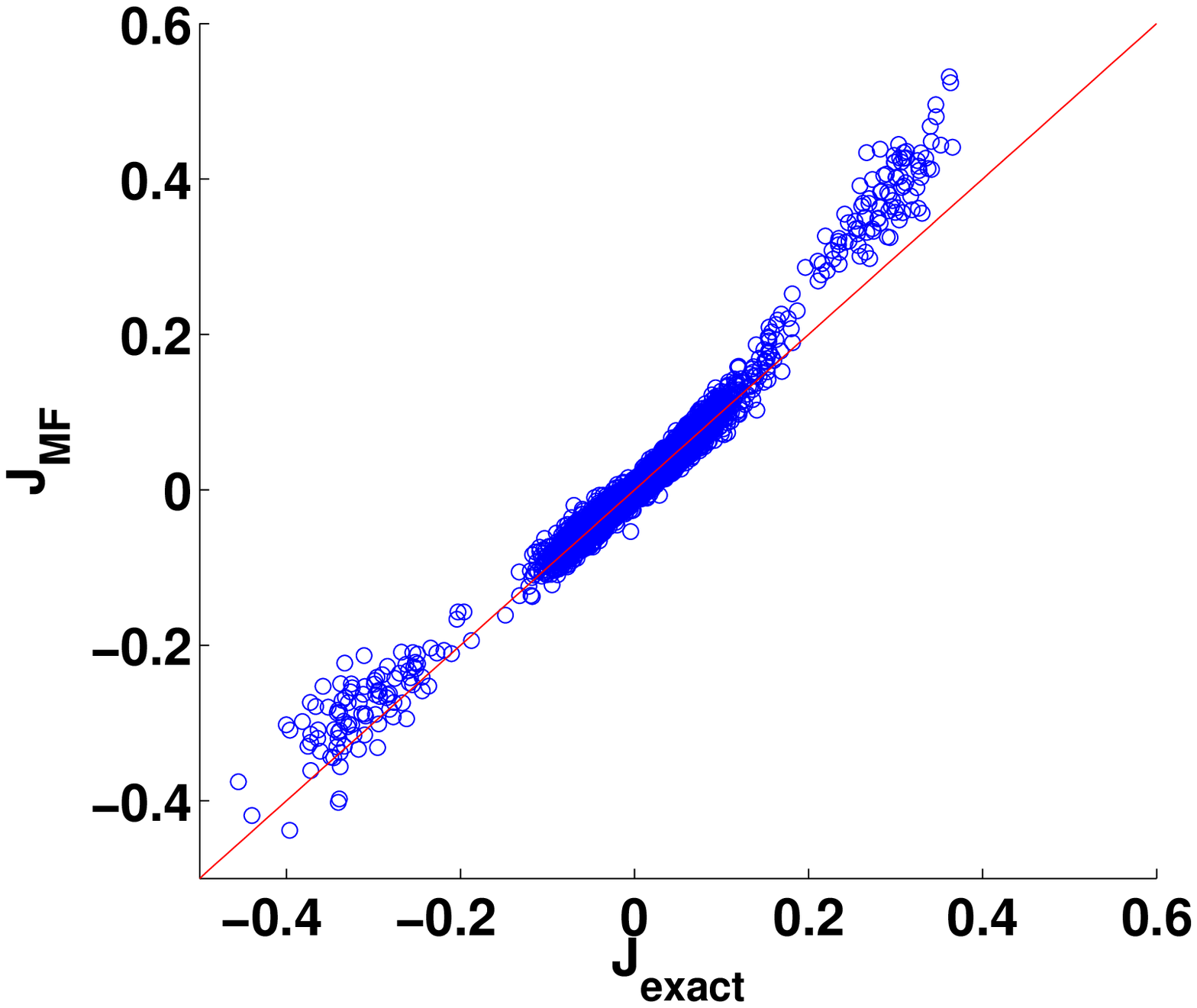}}
\subfigure{\includegraphics[height=1.4in, width=1.8in]{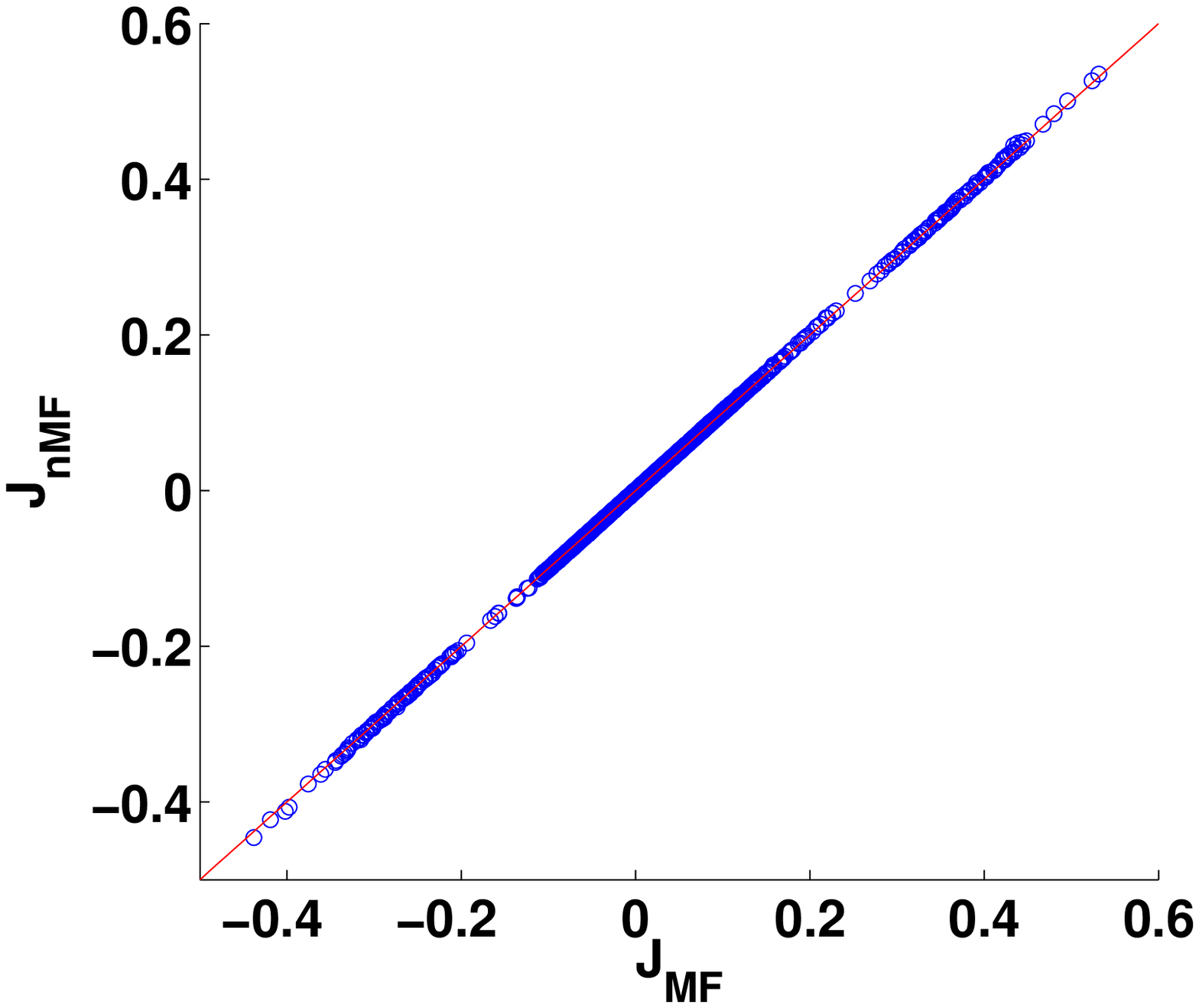}}
 
\caption{Comparison of couplings in nonstationary model inferred from cortical model data by exact 
and mean field algorithms: exact algorithm vs nMF, exact algorithm vs MF, and nMF vs MF.}
\label{Fig6}
\end{figure}

Next, we compare three different algorithms for the nonstationary model: exact, nMF, and MF. To
visualize the comparisons, we make scatter plots. We plot the couplings obtained by each of three
algorithms against each other, pairwise, in Figs.~\ref{Fig6} a-c.   The $J$s obtained by nMF and MF
show nearly perfect agreement with each other; thus, for estimating the couplings, there is nothing
to be gained from use the more time-consuming MF algorithm rather than the simpler nMF.   Both
mean-field algorithms give $J$s that generally agree quite well with those obtained by the exact
algorithm, although they tend to overestimate $J$s when they are large and positive and, to a small
degree, when they are large and negative.

As we did for the retinal data, we compared the mean-field approximations with the exact algorithm
on these data.  We found log likelihoods (with the Akaike penalty) for the nonstationary model with
$J$s of -0.30307 for the exact algorithm, -0.30981 for nMF, and -0.30409 for MF.  As was the case
for retinal data, MF is closer to the exact result than nMF.  However, the differences are around
2\% or smaller, so again we conclude that all model comparisons of interest can be made using
mean-field methods.

\subsection{Network graph identification}

\begin{figure}[hbtp] 
\centering
\subfigure{\includegraphics[height=2.4in, width=2.4in]{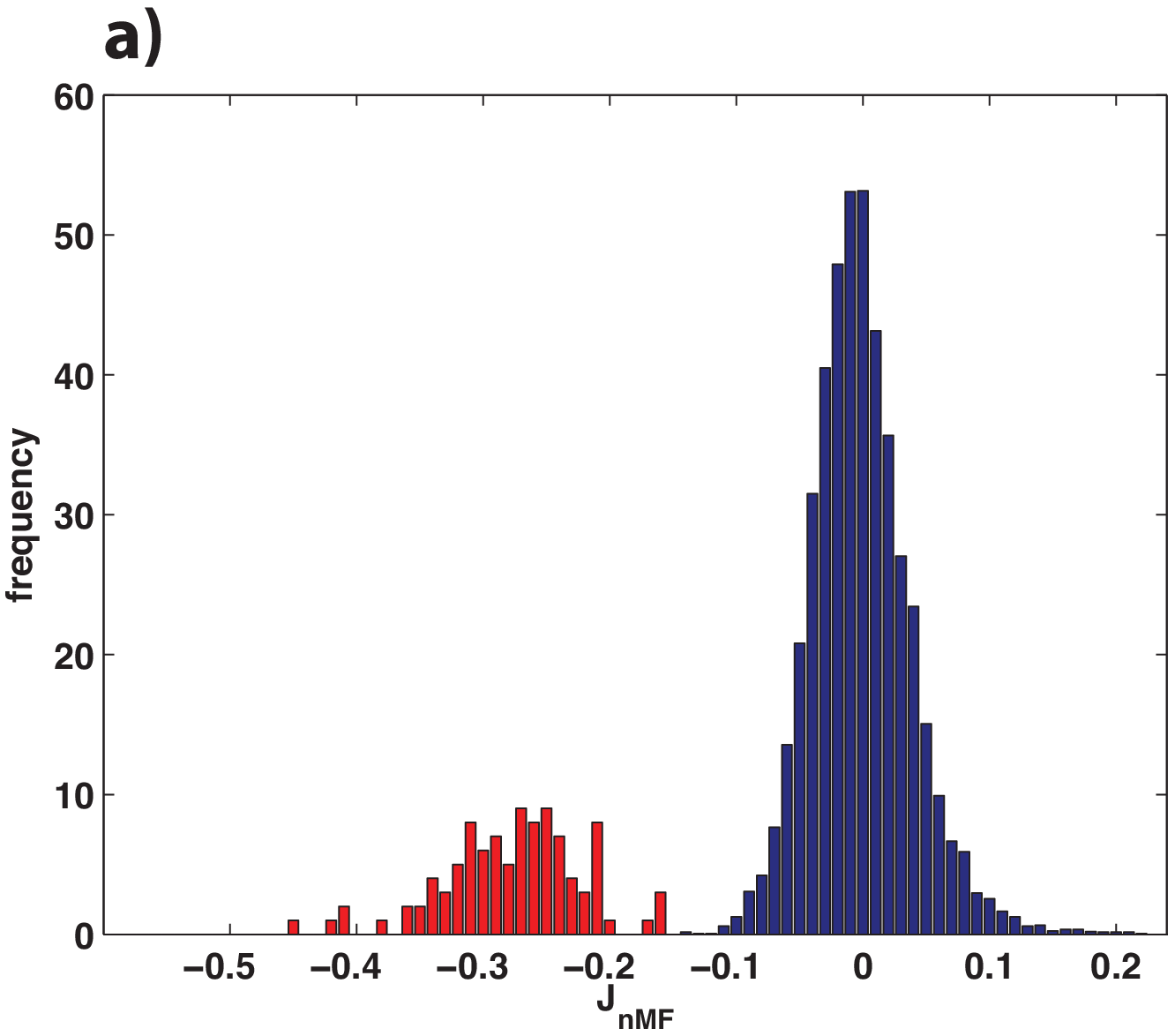}}
\subfigure{\includegraphics[height=2.4in, width=2.4in]{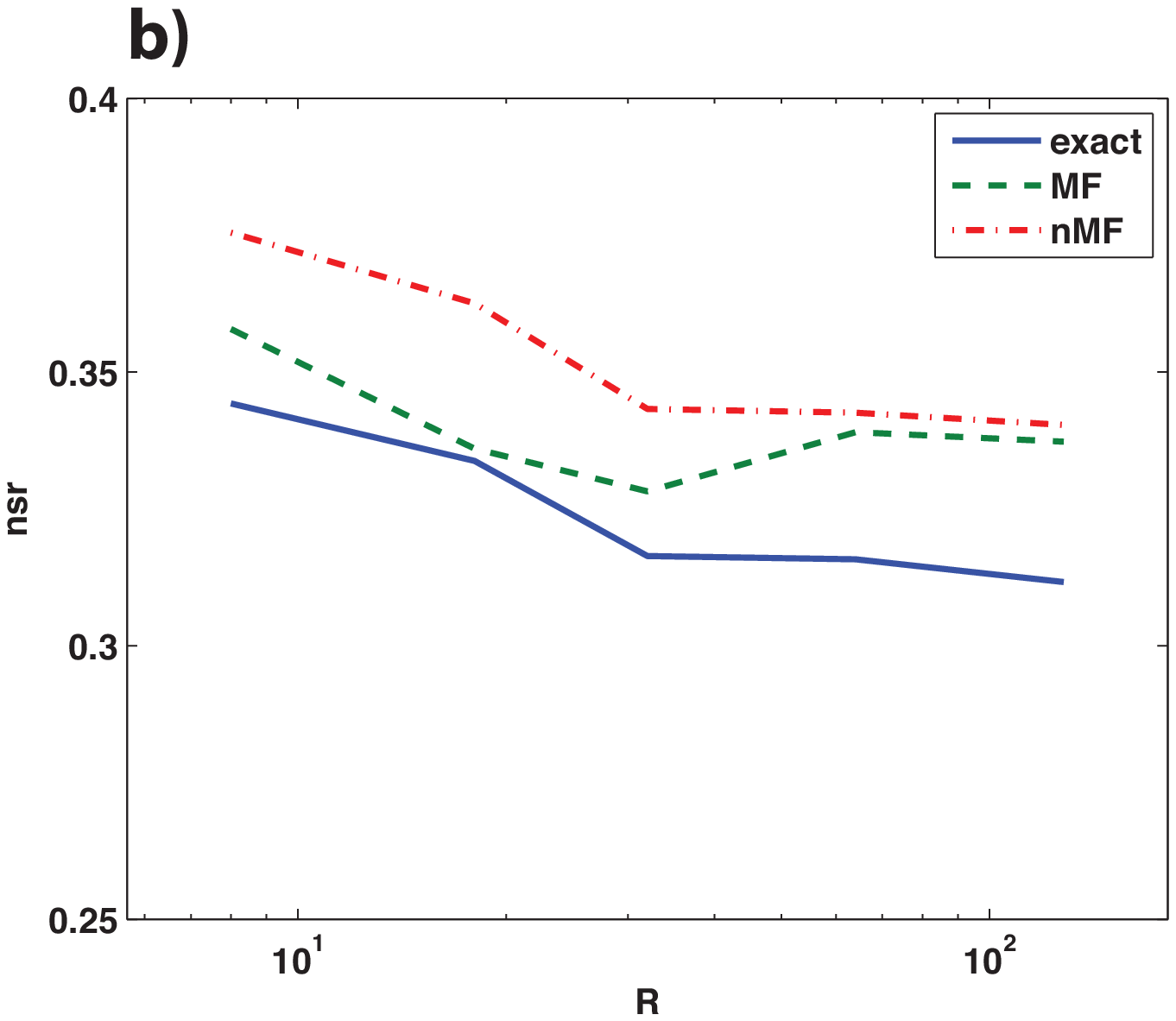}} 
\caption{ (a) Distributions of the couplings inferred from the
cortical network data using the exact nonstationary algorithm.  Red: inferred coupling values for
neuron pairs for which an inhibitory synapse is present in the cortical model network.   Blue:
inferred coupling values for pairs with no connection in the cortical model network. (b)
Noise/signal ratios $\zeta$ (Eq.~\ref{nsr}), as functions of number of stimulus repetitions, for
inhibitory couplings inferred from cortical model data using exact, nMF and MF algorithms. }
\label{Fig7} 
\end{figure}

The Akaike-penalized log likelihood is a suitable measure for comparing the quality of different
models, but it is not necessarily informative for identification of the connections present in the
network (i.e., the network graph).  Of course, how well a model performs on this task is possible
only when the true connections are known, which we do not in the case of the retinal data.  We
therefore examined histograms of the inferred $J$s for pairs of neurons that are connected and for
pairs that are unconnected  (Fig.~\ref{Fig7}a) in the cortical model.  The strong inhibitory
synapses are clearly identifiable as the peaks around $J \approx -0.3$.

There is not a qualitative difference between the exact and mean-field methods in how well they
identify the inhibitory part of the network graph. However, there are quantitative differences.
These are reflected in the values of the noise-signal ratio $\zeta$ defined in (\ref{nsr}). 
Fig.~\ref{Fig7}b  shows $\zeta$ as a function of data size (specifically,  the number of
repetitions), for the exact algorithm, nMF and MF.  One can see that $\zeta$ does not depend
strongly on the data size, but that MF is consistently somewhat better than nMF, though not as good
as the exact result.  We obtained similar results for this model in a tonic firing state using the
stationary algorithm \cite{HertzCNS10}.

The much weaker excitatory synapses are not clearly identified by any of the algorithms.  They are
overshadowed by the large peak around zero, which mostly represents the far more numerous absent
connections.  This illustrates how identifying synapses in a network correctly by fitting a model
such as this one depends on how strong they are.

\subsection{Frequency of synchronous spikes and spike patterns}

\begin{figure}[hbtp] \centering \subfigure{\includegraphics[height=2.5in,
width=2.5in]{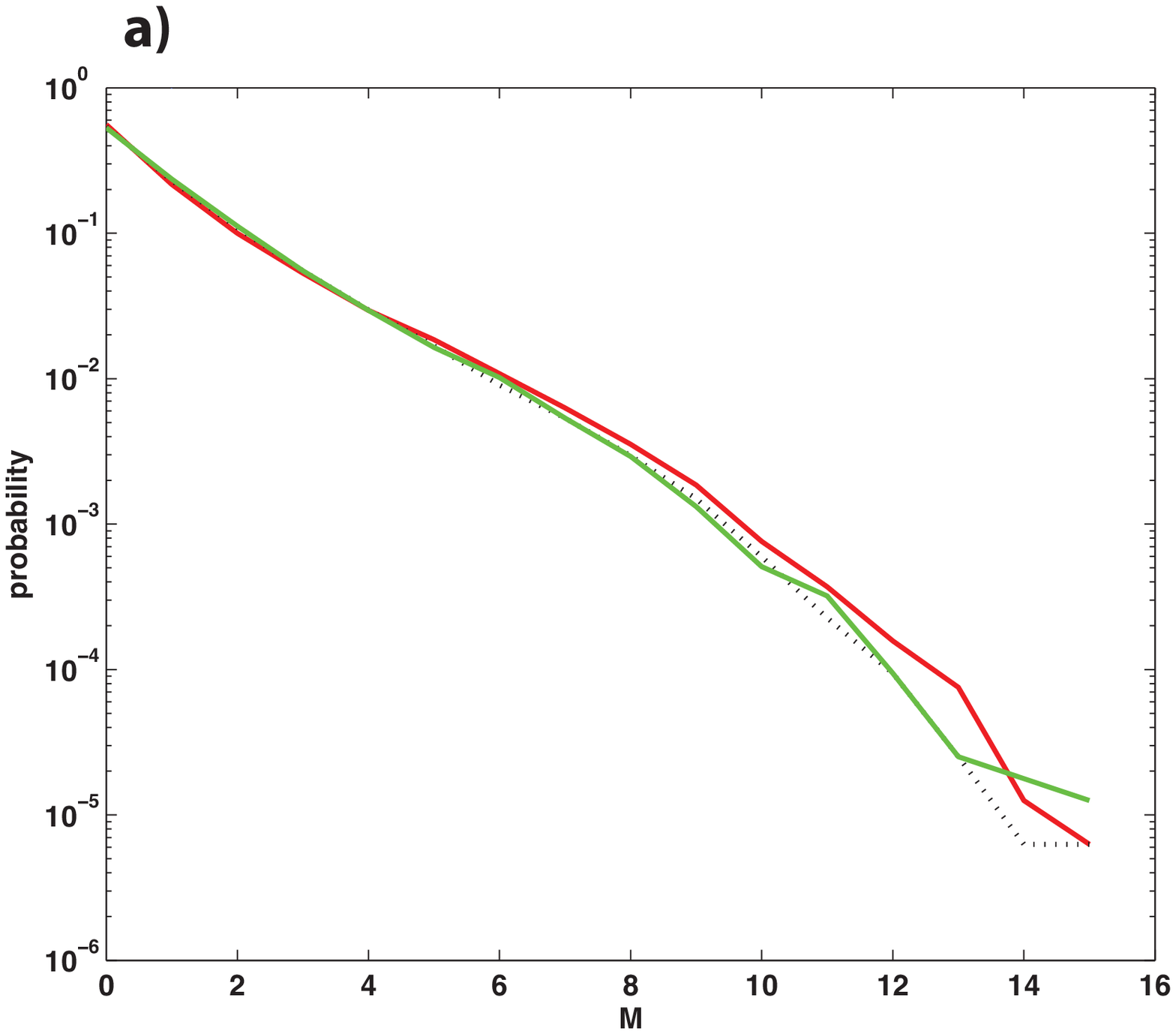}} \subfigure{\includegraphics[height=2.5in,
width=2.5in]{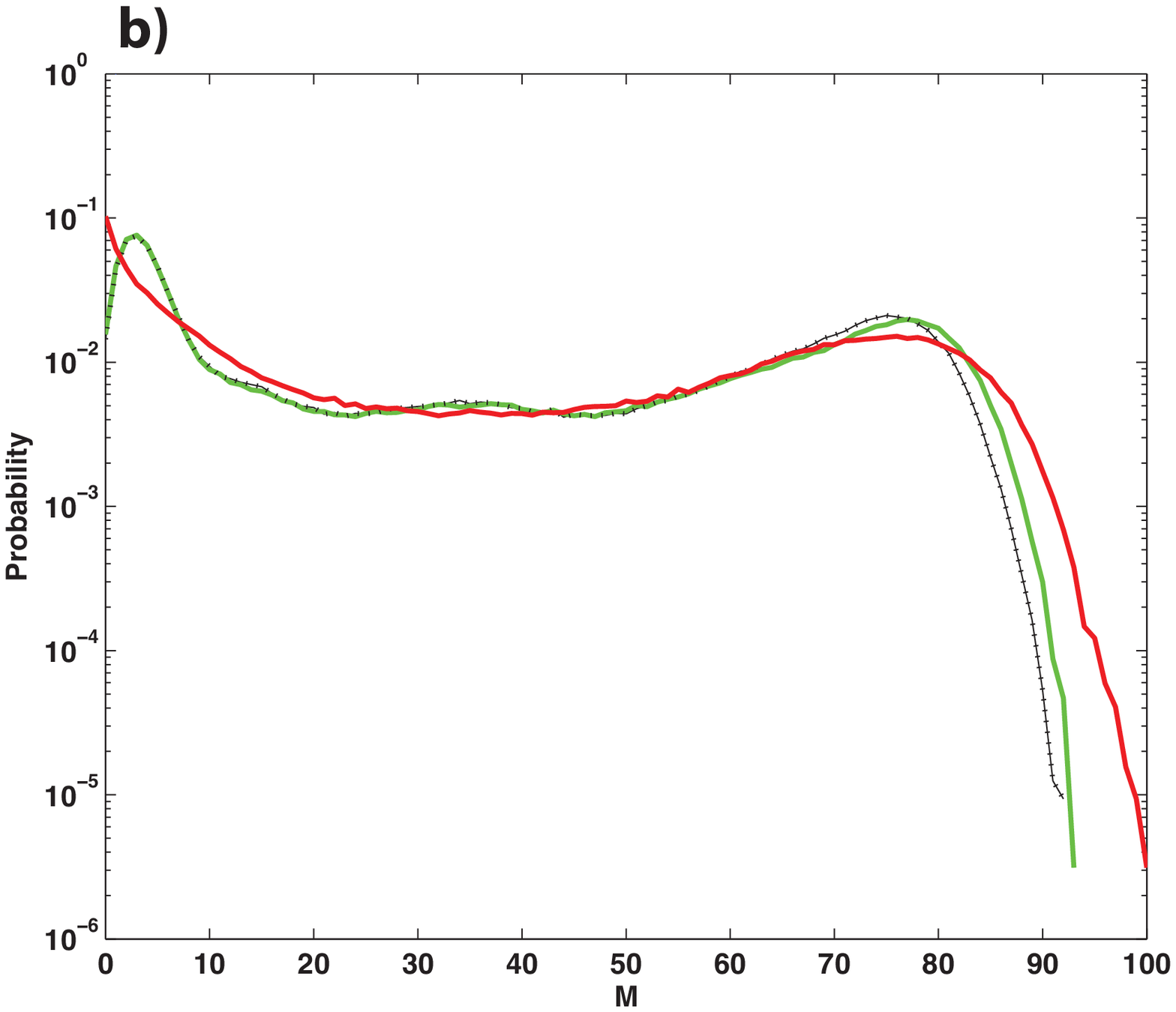}}\\ \subfigure{\includegraphics[height=2.5in,
width=2.5in]{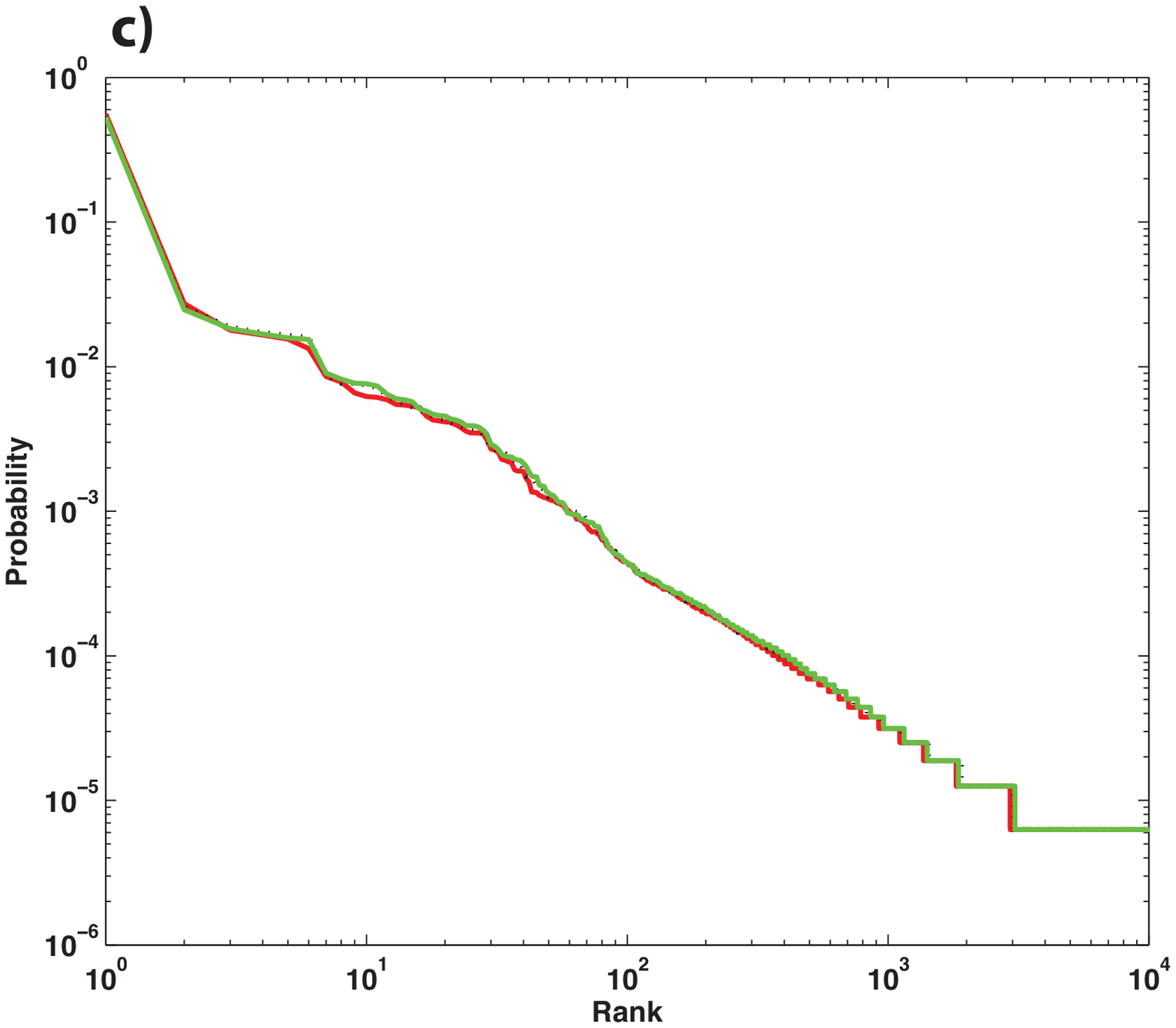}} \subfigure{\includegraphics[height=2.5in,
width=2.5in]{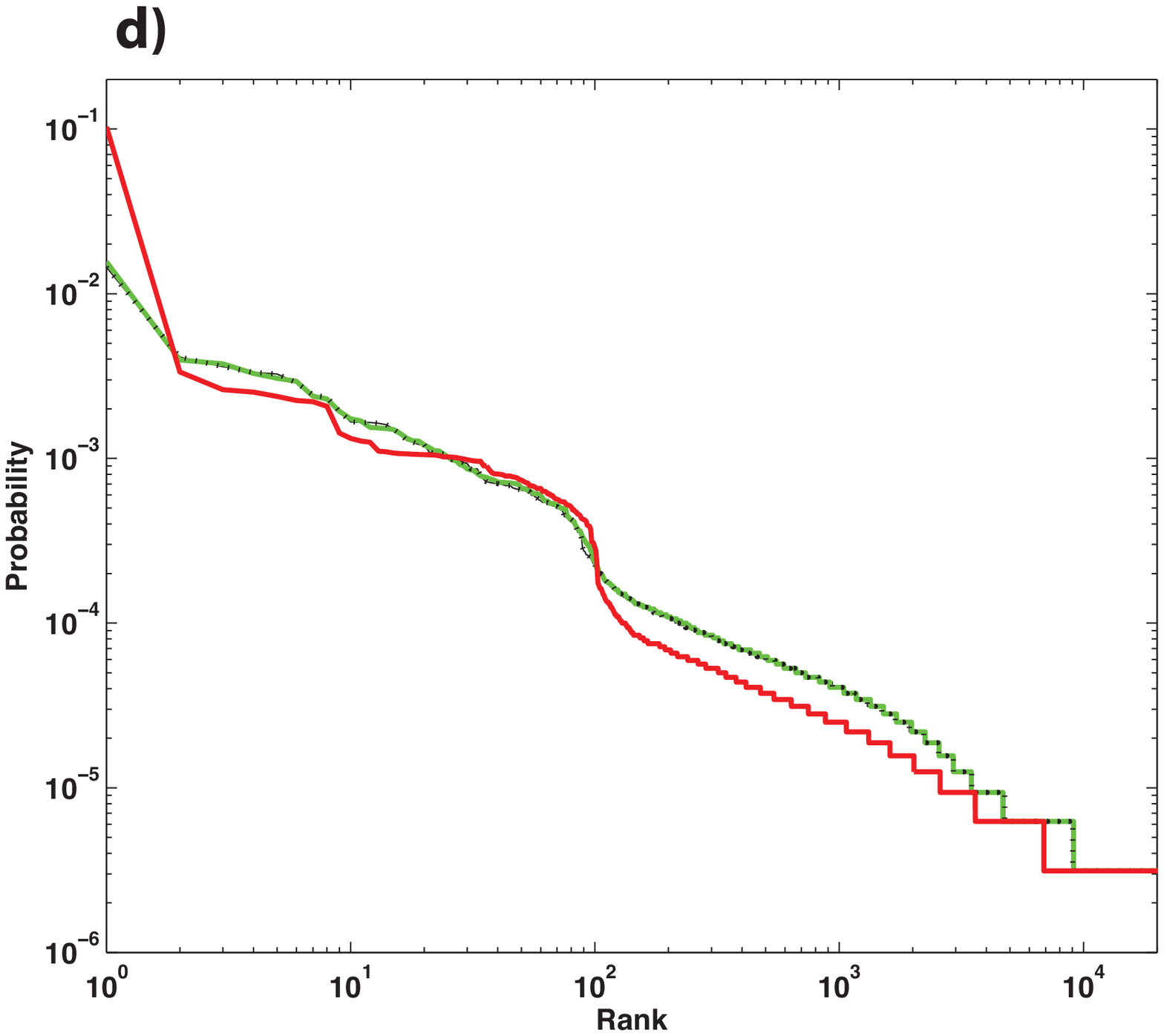}} \caption{(a) The probability of M synchronous spikes occurring in
20-ms time bins, 40 salamander retinal ganglion cells. Red: calculated from data. Black: calculated
from an independent nonstationary model. Green: calculated from a nonstationary model with
couplings. (b) The same as (a) but for the cortical model data. (c) The probability of spike
patterns from the 40 retinal ganglion cells as a function of their rank. (d) The same as (c) but for
the cortical model data. (Colour coding in (b-d) as in (a).)} \label{Fig8}
\end{figure}

Fig.\ \ref{Fig8}a shows the empirical distribution $P(M)$ (Eq.\ \ref{PofM}) of $M$ synchronous
spikes for the retinal data, as calculated directly from the data and from two different models:
a nonstationary independent-neuron model and a nonstationary model with couplings.  Evidently, 
the observed distribution of synchronous spikes can be modeled
very well without couplings provided that nonstationarity in the spike trains is taken into account,
that is, by a nonstationary independent model. As would be expected from the likelihood comparison
and lack of significance in the inferred couplings, adding couplings to the nonstationary
independent model does not change how well $P(M)$ for these neurons is predicted by the model.

The situation is different for the cortical model data. Fig.\ \ref{Fig8}b shows the probability of
synchronous events calculated from the data and from the same two models.  As would be anticipated
from the log-likelihood measures, the nonstationary model without couplings cannot reproduce the
pattern exhibited by the data, although it is far better than the stationary model without
couplings (not shown). Adding couplings to the nonstationary model improves the quality of the fit, but the
improvement is marginal. The fact that even the model with couplings cannot reproduce the shape of
the empirical curve exactly indicates that one should improve the model, possibly by taking into
account higher order correlations, or looking beyond one step in the past, to fully explain the
data.

Fig.\ \ref{Fig8}c shows that nonstationary models, with or without couplings, show an approximate
power law behavior in a Zipf plot, very similar to that exhibited by the data. The two nonstationary
models give nearly identical results (the two curves cannot be resolved in the plots) certainly
as good or better than the Gibbs equilibium fit \cite{Tkacik09}

The nonstationary models also both do a good job of reproducing the Zipf plot in the cortical model
data (Fig.\ \ref{Fig8}d).

\section{Discussion}

In neuronal networks, correlations in the external input can influence the apparent
correlations in firing of neurons. To understand neural information processing, it is thus
important to distinguish between aspects of neural firing that are simply inherited from
the external input and those that are generated by the network circuitry. This can be done
by considering statistical models that allow for non-stationary external input and do not
{\em a priori} assume stationary input. A simple example of such a model used in this
paper is the kinetic Ising model that can be efficiently fitted to neural data and other
point processes using exact and approximate inference methods.

The results reported in this paper show that it is possible, using the kinetic Ising
model, to infer interactions from systems exposed to nonstationary external input.  For
the retinal data, we find that for stationary models the inclusion of couplings improves
the fit qualitatively, as measured by the likelihood, but this is not the case for
nonstationary models.  Consistent with this, we also find that for nonstationary models
the connection strengths inferred from one half of the data are very poorly predicted by
those inferred from the other half.  Of course, nonstationary models have more (for the
present data, many more) parameters than stationary ones, so for limited data a stationary
model may outperform the corresponding nonstationary one.  However, for enough data, the
log-likelihood of the nonstationary model, with or without couplings, becomes
significantly larger than that of the stationary model even when the difference in the
number of model parameters is corrected for. For the cortical model data, on the other
hand,we find that the presence or absence of the couplings for both stationary and
nonstationary models make a significant difference in the likelihood of the models.
Furthermore, the inferred couplings using the nonstationary model are well correlated with
the real synaptic connections in the network.  In this case, again, the nonstationary
model is significantly better than the stationary model.

For the retinal data, we explored using smaller time bins, ranging down to 2 ms. 
Nonstationary models inferred from such data appeared to give worse fits than those using
20-ms bins.  However, interestingly, we found a few significant, positive, mostly
bidirectional couplings for the smallest bins (2 and 5 ms).   These couplings were not
apparent in the 20-ms-bin-based models.  They could be gap junctions or could represent
the effect of synchronized common input from retinal interneurons not recorded from or
included in the models.

For both the salamander retinal data and those from the local cortical network model, mean
field methods were found to give nearly the same log likelihoods as those found using the
exact algorithm.  Since the mean field methods are at least a couple orders of magnitude
faster than the exact calculations (which require up to 1000 iterations to converge for
these data), this finding suggests that these methods may make it possible to explore,
accurately and efficiently, model spaces much larger than ones that can be studied
practically using the exact algorithm.

Our results on using the models to reproduce the observed probability of synchronous
spikes largely parallel what we find by comparing the likelihoods. For the retinal data,
the observed behavior can be perfectly described by a nonstationary external field,
without requiring any couplings in the model. For the cortical data, on the other hand,
the nonstationary input model alone without couplings cannot describe the pattern that the
data shows for the frequency of synchronous spikes.  Adding couplings in this case
slightly helps but is not enough to explain the empirical results fully.

The empirical Zipf plots are also well described by the nonstationary independent model,
but in this case adding the couplings does not seem to improve the capacity of the model
to reproduce this feature.  The emergence of Zipf's law in the retinal data, has been
interpreted as indicating criticality in the network \cite{MoraBialek11} as indeed the
corresponding stationary Ising models seem to be poised close to a critical point. Our
results show that an alternative explanation is possible if nonstationarity is taken into
account, in terms of non-interacting spins. In this picture, clearly, Zipf's law has
nothing to do with the retinal circuitry.  Rather, it is a reflection of correlations in
the statistics of the external input generated by the natural scene stimuli used in the
experiment.

In their effort in modeling the same retinal data, Schneidman {\em et al}
\cite{Schneidman06} compared the equilibrium Ising distribution fitted to the mean and
pairwise (equal-time) correlations between neurons to the nonstationary model without
couplings (which they call the conditionally independent model). They concluded that the
equilibrium Ising distribution is more successful in modeling the data than the
nonstationary model without couplings.  However, their statement is based not on comparing
log-likelihoods but on how much of the so called multi-information is captured by the
models. Denoting the entropy of the spike patterns of $N$ neurons by $S_N$ and the entropy
of an independent neuron fit to it by $S_1$, multi-information is defined as
$I_N=S_1-S_N$.  Denoting the entropy of an equilibrium Ising fit to the data by $S_2$ and
that of a conditionally independent model by $S_{cond-int}$, Schneidman {\em et al} show
that $I_{(2)}/I_N=(S_2-S_N)/I_N$ is significantly larger than
$I_{cond-ind}/I_N=(S_{con-int}-S_N)/I_N$.  For the equilibrium Ising model, $S_2$ is equal
to the log-likelihood of the data under the model (what in this paper we use as our model
quality measure) and $I_{(2)}$ is equal to the Kullback-Leibler divergence between the
true distribution of spike patterns and the equilibrium Ising model fit to it. However,
for the conditionally independent model this is not the case. It is therefore difficult to
evaluate their statement in terms of log-likelihood.   We have estimated the log
likelihood (per neuron, per time bin) of the 20-ms-binned data under an equilibrium Ising
model as $-0.0926$ (see \ref{pestimation}).   (The Akaike penalty for this model,
$(N+1)/2T \approx 10^{-4}$ is negligible.)  This value is fairly close to what we found
for the stationary model ($-0.0887$) and consequently also well below that for the
nonstationary model (with or without couplings, $-0.0628$ and $-0.0643$, respectively).

The insignificance of the inferred connections in the retinal data does not mean that no
couplings, statistical or physiological, exist between the cells. Instead, it means that
the correlations, at the time scale and data length we had, do not reflect the effect of
such connections and are better modeled by external input. In other words, the
non-stationary model attributes correlated neuronal activity to correlations in the
external input at 20ms.  Little of the influence of direct interaction between neurons can
be seen with the length of data available to us. Likewise, we are not in a position to
draw definite conclusions on the significance of direct interactions detected for smaller
bin sizes, since the regularization we had to impose suppresses fluctuations in the inputs
over those frequencies. This issue could be resolved by longer recordings.  However, the
convergence of the Akaike-corrected and uncorrected log likelihoods in Fig.~\ref{Fig2}a
shows that there is little room to improve corrected log likelihood by much: Most of the
remaining failure to fit the data has to be blamed on the model.  The kinetic Ising model
is just a toy that we use here to illustrate the importance of nonstationary effects, but
it lacks many features that a good neural model should have.  The most obvious of these is
the fact that neurons can integrate their synaptic inputs over longer periods than just
one time step in our binned data. Generalized linear models (GLMs)
\cite{Truccolo05,Pillow08} include this feature in a general way.  In fact, our kinetic
Ising model is just a GLM without temporal integration kernels, and much of what we have
done here can be extended straightforwardly to the general case.  It also seems possible
to improve the modeling of synapses and to take into account the large number of neurons
in the network that are not recorded in the experiment.

However, even with better models, as we study larger and larger populations, longer and longer
recordings will be required.  An alternative (or parallel) approach to increasing the data length
would then be to better exploit the available data by including terms in the objective function
reflecting prior knowledge about the network connectivity or external input. For example, most
recorded neurons are not connected to each other, so we are effectively trying to reconstruct a
sparse network.  In this case, L-1 regularization, in which the log likelihood is penalized by a
term proportional to the sum of the absolute values of the $J$s, is a natural approach.

Pillow {\em et al} \cite{Pillow08} analyzed monkey retinal neuronal data using a GLM.
Unlike us, they found that the fit using a model with interactions between neurons was
significantly better than that obtained without interactions. This could be a species
difference. Alternatively, it might be because in the experiments which produced our data
the stimulus was simply so strong as to dominate the firing statistics and mask
interactions that might have been evident for weaker stimuli. There is also the difference
between our model and theirs that they knew the stimulus and represented its effect
through spatiotemporal inferred receptive fields, while we, not knowing it explicitly,
represent its effect through the parameters $h_i(t)$.  These parameters represent not only
the stimulus itself, but also all trial-to-trial reproducible input to the recorded
neurons from all the unrecorded neurons.  In the model of Pillow {\em et al}, on the other
hand, such correlated input is accounted for by couplings.  Therefore it is difficult to
compare their couplings with ours.

There is the further difference between the models that, although theirs used more
parameters to describe the interactions than ours, it did not require a number of
parameters that grew with the stimulus duration. It would be interesting to fit a model
like theirs to data like those we have analyzed here, but for which the movie pixel
intensities are known. This would allow us to determine to what degree the two species
really differ in their retinal circuitry and to what degree the two experiments just
differed in the strengths of their stimuli and/or the degree of correlated input to the
recorded set of neurons.

Although we did not find that interactions improved the fit quality significantly, we did
find an indication of a small number of significant effective couplings for 2- and
5-ms-bin data.   In this respect, there is a qualitative agreement between our findings
and those of Pillow {\em et al}.  A more relevant comparison is with the findings of 
Cocco {\em et al} \cite{Cocco09} on spontaneous salamander retinal ganglion cell data. 
Like us, they found significant positive, generally bidirectional couplings.  However, our
data on this timescale appear to be very noisy, so longer recordings would be needed to
study these couplings satisfactorily.

We cannot say with certainty, on the basis of the data analyzed alone, what the
biological meaning of the couplings inferred by any of the approaches discussed here is.
The couplings found by Cocco {\em et al} and by us for small time bins almost certainly do
not represent synapses.  Rather, since they are positive and generally bidirectional, they
could  be either gap junctions, an effect of correlated common input from other kinds of
cells in the retinal network, or external stimulus. It is not possible to distinguish these
possibilities from the present spike data alone.

Our nonstationary model is able, by averaging over trials, to isolate effects due to
reproducible correlated input from outside the recorded set of neurons, but it cannot do
so if that input varies randomly from trial to trial.  It ought to be possible to study
such effects by including ``hidden'' neurons in the model and inferring connections to,
from and among them in addition to those between neurons in the recorded population.

\section*{Acknowledgement}
We are most grateful to Michael Berry for providing us with the retinal data.

\appendix
\section{Estimating the likelihood of the equilibrium Ising model}
\label{pestimation}
Denoting the partition function of an equilibrium Ising model by $Z_{\rm Ising}$ and the
energy of a given configuration of spins, $\bss$,  by $E_{\rm Ising}(\bss)$, the
log-likelihood of the data is
\begin{equation}
{\cal L}_{\rm Ising}=\sum_{\bss} E_{\rm Ising}(\bss)-L \log Z_{\rm Ising},
\end{equation}
where $L$ is the number of samples. The difficult part in calculating this is estimating
the partition function. We considered two ways of making this estimate, as described
below.

Given an energy function $E_{\rm test}$ over the spin configuration, we have
\begin{eqnarray}
&&Z_{\rm Ising}= Z_{\rm test} \sum_{\bss} \frac{\exp(-E_{\rm test}(\bss))}{Z_{\rm test}} \exp(-E_{\rm Ising}(\bss)+E_{\rm test}(\bss))\\
&&\approx Z_{\rm test} \frac{1}{L} \sum^L_{l=1} \exp(-E_{\rm Ising}(\bss^l)+E_{\rm test}(\bss^l)) 
\end{eqnarray}
where $\bss^l$ are samples from the distribution with energy $E_{\rm test}$. Using this
equation with a properly chosen $E_{\rm test}$ for which $Z_{\rm test}$ can be calculated
easily thus allows estimating $Z_{\rm Ising}$ \cite{Bishop06PR}. Taking $E_{\rm test}$ as
an independent-neuron model with fields chosen to match the mean magnetizations of the
data, we generated samples of length $L=4\times 10^7$ from this distribution and used them
to estimate the Ising model partition function using the equation above.

We also generated samples from the Ising model itself using the Metropolis algorithm and considered the following estimate of the partition function:
\begin{equation}
Z_{\rm Ising}=\frac{1}{L} \sum^{L}_{l=1} \exp(-E_{\rm Ising}(\bss^l))/p(\bss^l),
\end{equation}
where $p(\bss^l)$ is the experimental probability of observing $\bss^l$ in the sequence generated from the metropolis algorithm.   In this case, too, we used $L=4\times 10^7$ samples. Both estimators yielded a log likelihood per neuron per time step of $-0.0926$ for the retinal data set with 20-ms bins. 

\section*{References}
\bibliographystyle{unsrt}
\bibliography{mybibliography2010}
\end{document}